\def\prd{Phys. Rev. D}
\def\prl{Phys. Rev. Lett.}
\def\apj{Astrophys. J.}
\def\apjl{Astrophys. J. Lett.}

\def\mnras{Mon. Not. R. Astr. Soc.}
\def\aap{Astr. Astrophys.}

\def\jcap{JCAP}

\def \<{\langle}
\def \>{\rangle}

\newcommand{\ra}{\;\raise1.0pt\hbox{$'$}\hskip-6pt\partial\;}
\newcommand{\lo}{\;\overline{\raise1.0pt\hbox{$'$}\hskip-6pt\partial}\;}

\newcommand{\Abs}{\abstract}
\newcommand{\Ack}{\acknowledgments}
\newcommand{\mktt}{\maketitle}

\documentclass[a4paper,11pt]{article}
\pdfoutput=1

\def\orcid#1{\kern .08em\href{https://orcid.org/#1}{\includegraphics[width=1.0em]{orcid.jpg}}}

\usepackage{jcappub,graphicx,epsfig,natbib,color,times,bm,amsmath,multirow,amssymb,mathtools,mathrsfs,extarrows}
\usepackage{xcolor, academicons, booktabs}
\usepackage[ddmmyyyy,hhmmss]{datetime}

\begin{document}

\title{Optimal map-making with singularities}

\author[b,c]{Zirui Zhang}

\author[b,d]{, Yiwen Wu}

\author[b]{, Yang Liu}

\author[b]{, Siyu Li}

\author[b]{, Hong Li}

\author[a,b]{ and Hao Liu$^{\ast,}$}\emailAdd{ustc\_liuhao@163.com}

\affiliation[a]{School of Physics and optoelectronics engineering, Anhui University, 111 Jiulong Road, Hefei, Anhui, China 230601.}

\affiliation[b]{Key Laboratory of Particle and Astrophysics, Institute of High Energy Physics, CAS, 19B YuQuan Road, Beijing, China, 100049.}

\affiliation[c]{Institute of Frontier and Interdisciplinary Science and Key Laboratory of Particle Physics and Particle Irradiation (MOE), Shandong University, Qingdao}

\affiliation[d]{University of Chinese Academy of Sciences, Beijing, China, 100086}

\Abs{

In this work, we investigate the optimal map-making technique for the
linear system $\bm{d}=\bm{A}\bm{x}+\bm{n}$ while carefully taking into
account singularities that may come from either the covariance matrix
$\bm{C} = \<\bm{n}\bm{n}^t\>$ or the main matrix $\bm{A}$. We first
describe the general optimal solution, which is quite complex, and then
use the modified pseudo inverse to create a near-optimal solution,
which is simple, robust, and can significantly alleviate the unwanted
noise amplification during map-making. The effectiveness of the nearly
optimal solution is then compared to that of the naive co-adding
solution and the standard pseudo inverse solution, showing noticeable
improvements. Interestingly, all one needs to get the near-optimal
solution with singularity is just a tiny change to the classical
solution, which is designed for the case without singularity.

}

\mktt

\section{Introduction}
\label{sec:intro}

The precise measurement of the temperature and polarization of the
Cosmic Microwave Background (CMB) has allowed us to establish the
standard cosmological model ($\Lambda$CDM). The theory of inflation was
developed in the 1980s to address the flatness, horizon, and monopole
concerns within the $\Lambda$CDM framework~\cite{Guth1981, Linde1982}.
Despite inflationary cosmology's enormous success, the expected
primordial gravitational waves (PGW) have not been discovered yet.
Detection of PGW and testing inflationary cosmology remain critical
parts of current and future CMB researches. The most promising technique
to detect the PGW is via the CMB B-mode polarization observations,
according to~\cite{Seljak:1996ti, Kamionkowski:1996zd, Seljak:1996gy}.

With the completion of the Planck mission~\cite{Planck:2018nkj}, the
fundamental scientific goals of various present and future CMB studies
are set towards detecting the CMB B-mode polarization. These include the
ground-based experiments like BICEP/Keck~\cite{Bicep2014, BK2015},
SPTpol~\cite{SPTpol2020}, CLASS\cite{CLASS2022ApJ...926...33D},
ACTpol~\cite{ACTpol2020}, POLARBEAR~\cite{Polarbear2017}, and
AliCPT~\cite{AliCPT2017science, AliCPT2020receiver, Ghosh:2022mje} in
the Northern Hemisphere; as well as the balloon experiments like
EBEX~\cite{Oxley:2004yxi}, SPIDER~\cite{SPIDER:2021ncy} and future
experiments like SO~\cite{SimonsObservatory:2018koc},
CMB-S4~\cite{CMB-S4:2016ple}, and the space mission
LiteBIRD~\cite{LiteBIRD:2020khw}, which is scheduled to be launched in
2029.

To obtain a more useful CMB polarization dateset, the time-ordered data
(TOD) is often compressed into sky maps, and the data volume is reduced
from tens of trillions to tens of millions, which is called a mapmaking
approach. This kind of approach seeks to condense the data while
maintaining as much cosmological information as possible. Typically,
mapmaking is treated as a linear problem with the goal of generating an
unbiased estimate of the CMB sky map while minimizing its variance. This
is usually accomplished using approaches such as the minimum variance or
maximum likelihood estimation, e.g.,~\cite{1997ApJ...480L..87T,
2001PhRvD..65b2003S, 2009MNRAS.393..894S, 2014A&A...572A..39S,
2018arXiv180108937P, 2017A&A...600A..60P, 2022A&C....3900576E}.

However, in the case of real TOD, a number of time-domain filtering
processes are frequently required to remove the atmosphere and ground
radiation, as well as numerous systematics. Although these operations
are necessary, they have an inevitable side effect of removing a
considerable percentage of the desirable signal at the same time. As a
consequence, some of the signal modes are permanently lost due to the
filtering and, if the filtering consists of operations in multiple
domains such as Fourier and polynomial, some of the signal modes may be
suppressed rather than removed, leading to a reduced signal-to-noise
ratio (SNR) and an amplified noise in the final sky map.

For a naive map-making algorithm like co-adding and averaging, the above
mentioned issues are not big problems, because it does not pay much
attention to further improving the SNR in the final sky map product --
which is done by optimized map-making algorithms. Unfortunately, most
optimized map-making algorithms require to use the inversion of the
covariance matrix, which is problematic if singularities are involved.
To the best of our knowledge, there is still no comprehensive discussion
regarding a rigorous optimal solution of map-making with the presence of
various singularity problems, despite the fact that there are some
applicable solutions to this issue with certain compromises, such as
using a modest addition of human choice to make the covariance matrix
non-singular.

In this work, we will explore a rigorous optimal map-making method that
takes into account various singularity concerns. We will also show how
to simplify the optimal solution while tightly limiting any unintended
side effects along with the simplification. The final recommended
solution is simple, near-optimal, and can significantly alleviate the
above mentioned noise amplification effect.

The outline of this work is the following: We first introduce the
optimal and near-optimal solutions in section~\ref{sec:metho}, with the
corresponding lengthy mathematical details put in appendix~\ref{app:math
details}; then we discuss some technical details of the solution in
section~\ref{sec: tech details}, followed by the code validation and
simulation tests in section~\ref{sec: validation simulation test}.
Finally, we give a conclusion and some further discussions in
section~\ref{sec:disscuss}.

\section{Methodology}\label{sec:metho}

Because the mathematical deduction of the optimal solution of
map-making with singularity is quite lengthy, we provide only a brief
introduction in this section, and put the main contents in
appendix~\ref{app:math details}.

\subsection{The classical minimum variance solution and its limitations}
\label{sub: old method}

The problem starts from the well-known matrix equation that connect the
pixel domain sky map $\bm{x}$ to the time-order data $\bm{d}$ (TOD) with
noise $\bm{n}$:
\begin{align}\label{equ:basic form}
\bm{d} = \bm{A}\bm{x}+\bm{n},
\end{align} 
where $\bm{x}$ is a column vector of size $n_{\rm{pix}}$; $\bm{n}$ is
another column vector of size $n_{\rm tod}$; and $\bm{A}$ is a matrix
with $n_{\rm tod}$ rows and $n_{\rm{pix}}$ columns. In order to
increase the signal to noise ratio (SNR), $n_{\rm tod}$ is usually much
bigger than $n_{\rm{pix}}$, so each sky pixel is observed many time.
The goal of map-making is to find the best solution of $\bm{x}$ from
$\bm{d}$, and the simplest case of the solution requires the following:
\begin{enumerate}
\item\label{itm:1} Each row of $\bm{A}$ contains only one element that
is equal to 1, and all other elements are zero. Thus,
$\bm{A}^t\bm{A}=\bm{N}_{obs}$ is diagonal\footnote{Otherwise at least
one row will contain more than one element equal to 1} and
contains the number of observations at each sky pixel.

\item\label{itm:2} $\bm{A}$ has full column-rank and $\bm{A}^t\bm{A}$
is invertible.

\item\label{itm:3} The noise covariance matrix $\bm{C} = \<
\bm{n}\bm{n}^t \>$ is invertible and fully represents the noise
properties.
\end{enumerate} 
Given the fulfillment of these conditions, the optimal solution
$\widetilde{\bm{x}}$ for the sky signal is given by the equation below,
which has been documented in various publications,
e.g.,~\citep{Lupton1993, 1997PhRvD..56.4514T}:
\begin{align}\label{equ:solution simplest case}
\widetilde{\bm{x}} =
(\bm{A}^t\bm{C}^{-1}\bm{A})^{-1}\bm{A}^t\bm{C}^{-1}\bm{d}.
\end{align}
Certainly, in instances where the aforementioned conditions are not
satisfied, particularly regarding items~\ref{itm:2} and~\ref{itm:3},
the above solution is no longer applicable. Especially, when a
singularity exists in either or both of $\bm{A}^t\bm{A}$ and $\bm{C}$,
the true optimal solution can be complicated, which has not been
documented in detail before.

\subsection{The general minimum variance solution with singularity}
\label{sub: general method}

In this work, we give the optimal map-making solution in general, which
requires only two fundamental conditions that are almost always true:
\begin{enumerate}
\item $n_{\rm tod}\gg n_{\rm{pix}} $, so each sky pixel is observed many
times.
\item The noise covariance matrix $\bm{C} = \< \bm{n}\bm{n}^t \>$ does
converge.
\end{enumerate}
Apparently, such a significant relaxation of the conditions means
$\bm{A}^t\bm{A}$ and $\bm{C}$ are both allowed to be singular, and no
constraint is assumed for the relationship between $\bm{A}$ and
$\bm{C}$. A lengthy introduction of how to get the general optimal
solution under the aforementioned conditions can be found in
appendix~\ref{app:math details}. In summary, the mathematical inference
yields two key findings:

First of all, it is possible to obtain a general true optimal solution,
which requires using the singular value decomposition to carefully
analyze the linear system and the origin of singularity, in order to
clearly separate the singularity that arise from various origins, and
use all available information properly to obtain the final optimal
solution. Details of the true optimal solution can be found in
appendix~\ref{sub:deal with singular cov} --~\ref{sub: main matrix
singularity}, and the main process includes first dealing with the
covariance matrix's singularity, and then treating the main matrix's
singularity properly, to obtain the final optimal solution by
eq.~(\ref{equ:ss7}).

Secondly, since the true optimal solution is complicated and uneasy to
use, we have designed a carefully simplified solution based on the
modified pseudo inverse, as introduced in appendix~\ref{subsub:modified
pseudo inverse} and particularly, in eq~(\ref{equ:afaiusdf878321}). The
simplified solution requires only a tiny change to the classical pseudo
inverse solution, but has three major advantages: A) It is much easier
to use, because the solution is contained in a single equation. B) The
side effect due to simplification is strictly limited to the singular
part, which is usually negligible (see appendix~\ref{sub:summary of two
solutions} for more discussions). C) The solution can significantly
alleviate the unwanted noise amplifications (see
section~\ref{sub:regarding the main scan} for more details).

\section{A few discussions about the technical details}
\label{sec: tech details}

In this section, we discuss a few technical details, which do not affect
the theoretical aspects of the solution, but may be useful for pursuing
the best effect in an implementation of the method.

\subsection{ The noise amplification effect in optimal mapmaking }
\label{sub:regarding the main scan}

Another strong reason to adopt the near optimal solution in
eq.~(\ref{equ:afaiusdf878321}) is to prevent the noise amplification
effect in optimal mapmaking, which is tightly associated with the
overall scan strategy (structure of $\bm{A}$) and the properties of
filtering. Below we discuss this phenomenon and explain why the near
optimal solution helps enormously to prevent such an unwanted effect.

In the case where matrix $\bm{A}$ possesses full column rank, the
matrix $\bm{A}^t \bm{C}^{\times} \bm{A}$ always remains non-singular
and is safe to use; however, the behavior of matrix $\bm{A}^t\bm{C}^{+}
\bm{A}$ (compute the inversion of $\bm{C}$ with the standard pseudo
inverse) is impacted by the structure of matrix $\bm{A}$. For example,
consider a special case where celestial observations are undertaken in
a peculiar manner such that all pixels sharing the same signal value
are put into one group, and the process of filtering consists merely of
excluding the average value of each group. Apparently, this kind of
strategy will remove the signal completely, whereas noise continues to
prevail, leading to a zero SNR for all pixel domain modes after
filtering.

Certainly, the above example will not happen in reality, but it does
tell us that, the joint effect of scan strategy and TOD filtering can
possibly cause a deteriorated SNR. Meanwhile, the minimum variance
solution will try to maintain signal integrity as far as possible.
Therefore, if the filtering causes a significant SNR deterioration at a
particular pixel domain mode (provided it doesn't plummet to zero), the
minimum variance solution will automatically amplify the total signal to
keep the desired component lossless, which leads to an inevitable noise
amplification in this mode.

The above mentioned noise amplification effect is characterized by the
small eigenvalues of $\bm{A}^t \bm{C}^{+} \bm{A}$ (\emph{not} the
eigenvalues of $\bm{C}$). Therefore, during the computation of $\bm{A}^t
\bm{C}^{+} \bm{A}$'s pseudo-inverse, it becomes essential to choose an
appropriate cutoff of the small eigenvalues to exclude the components
with poor SNR after filtering. An alternative way to solve this problem
is to apply a posterior Wiener filter to optimally assign weights to the
components according to their SNR. However, as a well known effect, the
Wiener filtered signal is \emph{no longer} lossless.

On the contrary, although the same problem may also exist in the
inversion of $\bm{A}^t \bm{C}^{\times} \bm{A}$ (the solution with
modified pseudo inverse), the chance to get a very small eigenvalue of
$\bm{A}^t \bm{C}^{\times} \bm{A}$ is greatly reduced, because we have
eliminated all zero modes of $\bm{C}$ via the modified pseudo inverse,
which makes the solution with modified pseudo inverse much more robust.

On further preventing the noise amplification effect, the pointing
matrix $\bm{A}$ should, in the ideal case, be composed of random
observations. This randomness ensures that the power loss caused by
filtering is dispersed evenly across different pixel domain modes,
preventing any mode from experiencing a severe SNR decrease, and hence
eliminates the aforementioned problem. Although implementing such a
random scanning scheme is impractical, it is still helpful to add
multiple modes to the scan strategy to alleviate the noise amplification
effect, and to make the strategy more robust. Meanwhile, it is crucial 
to take $\bm{A}^t \bm{C}^+ \bm{A}$'s eigenvalue threshold into account 
whenever one needs to deal with the $\bm{S}_+$ solution.

\subsection{Approximation of the noise covariance matrix}
\label{sub:approx of noise cov-mat}

In order to achieve the optimal solution, it is desirable to possess
knowledge of the noise covariance matrix. Nonetheless, acquiring the
true noise covariance matrix presents considerable challenges,
particularly in the context of ground-based experiments, where the noise
is significantly influenced by atmospheric conditions that vary over
time. Consequently, it becomes necessary to consider an approximation of
the noise covariance matrix.

Indeed, the simplest approach of map-making involves estimating the
signal of a single sky pixel by taking the average of all observations
corresponding to that pixel. This method in fact assumes a noise
covariance matrix with identical diagonal elements, representing a
rather basic and unsophisticated approximation. A more refined technique
involves estimating the Fourier spectrum of the noise over a specific
time period, thereby incorporating the two-point covariance of the
noise. Utilizing a longer time period is generally advantageous as it
allows for a greater number of samples of the two-point covariance and a
wider range of correlation lengths. In essence, the improvement scales
approximately as the square root of the number of points, denoted as
$N_p$, used in the time segment to estimate the Fourier spectrum. Thus,
the aforementioned unsophisticated approximation corresponds to setting
$N_p$ equal to 1. By considering $N_p$ to be the number of TOD points
within a given time interval (e.g., a few minutes or thousands of
points), we already achieve a substantial enhancement in the estimation
of the noise covariance matrix. Further improvements obtained by
considering hours of TOD are likely to yield only a marginal enhancement
to the estimation of the noise covariance matrix.

In a more comprehensive approach, it is advisable to incorporate the
atmospheric emission model while also accounting for the temporal
variability of atmospheric emissions and the spatial distribution of
such emissions in the local zenith coordinates. However, it should be
noted that these considerations extend beyond the scope of this work.

A fundamental aspect to consider when approximating the noise covariance
matrix is the requirement that the resultant estimation matrix $\bm{M}$,
leading to $\widetilde{\bm{x}}=\bm{M}\bm{d}$, should not possess
excessively large singular values. The presence of such large singular
values can lead to substantial errors when they happen to interact with
the uncertainty inherent in the noise covariance matrix.

\section{Code validation, simulation and tests}
\label{sec: validation simulation test}

In ground-based CMB experiments, the methods for converting TOD into
maps primarily involve naive map-making, as used by
BICEP~\cite{BICEP2:2014owc, BicepKeck:2021ybl},
POLARBEAR~\cite{POLARBEAR:2017beh}, and SPT~\cite{Millea:2020iuw}); and
also maximum likelihood map-making, as employed by
ACT~\cite{Dunner:2012vp, ACT:2020gnv} and
POLARBEAR~\cite{2017A&A...600A..60P}). While naive map-making is prone
to producing imperfect estimates of the sky signal, it can be improved
by fine-tuning the filters to match the data property, which enables a
more diagonal time-domain noise covariance matrix after filtering. Next,
signal-only simulations can be used to correct the angular power
spectrum suppression cause by filtering~\cite{BICEP2:2014owc,
POLARBEAR:2017beh, Millea:2020iuw}. On the other hand, maximum
likelihood map-making, particularly as used by
POLARBEAR~\cite{2017A&A...600A..60P}, produces maps with a better
signal-to-noise ratio by simultaneously solving for the intended signal
and contamination templates, and ACT processes the filtered TOD
similarly to unfiltered data~\cite{Dunner:2012vp}, treating filters as
having little effect on the sky signal.

Both the naive and maximum likelihood map-making algorithms will be
tested in this section, and their performance will be compared to the
recommended modified pseudo inverse solution, which is introduced in
section~\ref{sec:metho} and detailed in appendix~\ref{subsub:modified
pseudo inverse}. For convenience, we adopt the following symbols: the
maximum likelihood map-making (implemented by the pseudo inverse) is
indicated as $\bm{S}_{+}$, the modified pseudo inverse solution is
represented as $\bm{S}_{\times}$, and the simplest naive map-making is
denoted as $\bm{S}_c$. The following can be used to express these three
solutions:
\begin{align}
	\bm{S}_{\rm c} &= (\bm{A}^t\bm{A})^{-1}\bm{A}^t\bm{d}\\ \nonumber
    \bm{S}_{+} &= (\bm{A}^t\bm{C}^+\bm{A})^+\bm{A}^t\bm{C}^+\bm{d} \\ \nonumber
    \bm{S}_{\times} &= (\bm{A}^t\bm{C}^\times\bm{A})^+\bm{A}^t\bm{C}^\times\bm{d} 
\end{align}
Here, the matrix $\bm{A}$ is determined by the scanning strategies, and
we assume that $\bm{A}^t\bm{A}$ is non-singular, as is typical in most
experiments. Our simulation is based on the observation strategies
proposed by AliCPT~\cite{AliCPT2017science}, which include two scanning
strategies: a large-area scan for the Milky Way observation, and a
small-area scan for the CMB observation. Both scanning strategies made
use of the Constant Elevation Scan (CES) mode to guarantee a stable
atmospheric payload on detectors. 

A code validation was done prior to the simulation, as introduced in
Appendix~\ref{appendix:validation}. Then, sections~\ref{subsec:large}
and \ref{subsec:small} below presents the simulation tests with the
above two scan strategies.

\subsection{The performance tests: for a large sky region} \label{subsec:large}

\begin{figure}[!ht]
    \centering
    \includegraphics[width=0.45\textwidth]{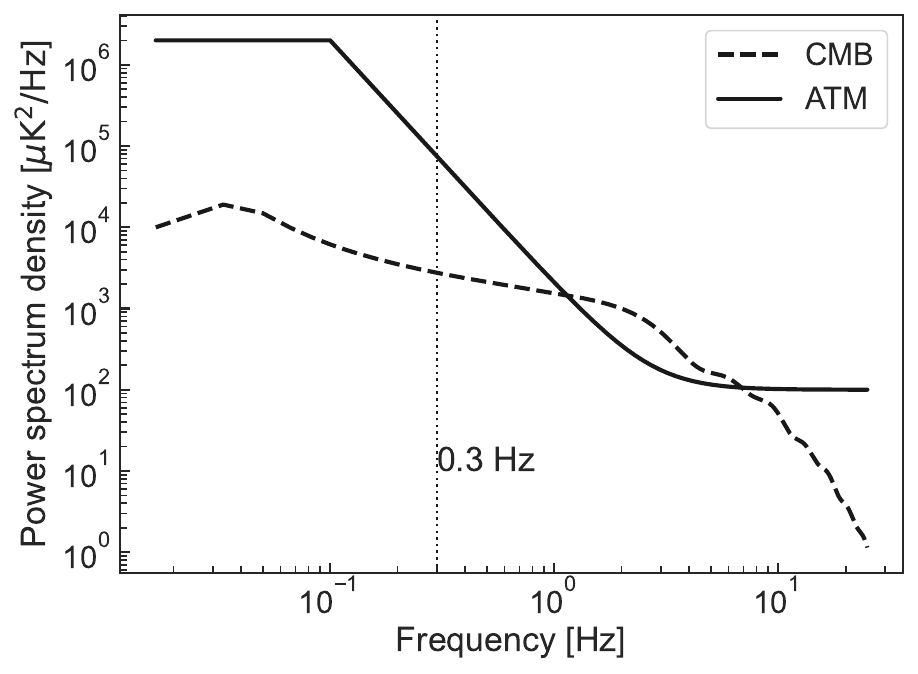}
    \caption{The power spectrum density for both CMB and atmospheric
    noise. The vertical doted line shows the threshold frequency. For
    the singular covariance, the power spectrum is set to be 0 at
    frequencies less than the threshold.}
    \label{fig:PSD_CMB_ATM}
\end{figure}

For the large-area scan simulation, the noise covariance matrix is
assumed to be diagonal in the Fourier domain, and its diagonal
components are represented by the noise's Fourier spectrum. Thus, the
time domain noise covariance matrix is $\bm{C} =
\bm{W}\bm{\lambda}^2\bm{W}^H$, where $\bm{W}$ and $\bm{W}^H$ signify
the Fourier transformation matrix and its conjugate transpose
respectively, and each column of $\bm{W}$ representing one Fourier
mode. The simulated TOD includes the noise computed from $\bm{C}$ and
the CMB signal constructed from the Planck 2018 best-fit cosmological
parameters~\cite{2020A&A...641A...6P}. According to the Kolmogorov
model of turbulence, the power-law index of the integrated atmospheric
emission (noise) varies between $-11/3$ and
$-8/3$~\cite{1995MNRAS.272..551C}; thus, its Fourier spectrum is
modeled as
\begin{align}
    P(\nu)=\sigma_0^2\left(1+\left(\frac{f}{f_{\rm knee}}\right)^{\alpha}\right),
\end{align}
wherein $\nu$ represents the frequency while $\sigma_0^2$ is the
variance of white noise, and $f_{\rm{knee}}$ is the knee frequency.
According to the Kolmogorov model of turbulence, the power-law index of
the integrated atmospheric emission (noise) varies between $-11/3$ and
$-8/3$, thus here we set $\alpha = -3$. For other parameters, we
considered an equivalent detector of which the noise level is same to
the noise level that is averaged over all detectors. Thus, we set the
amplitude of white noise $\sigma_0^2=10^2~\mu{\rm K}^2~{\rm Hz}^{-1}$
while the knee frequency $f_{\rm knee} = 2.15~{\rm Hz}$. When the
frequency is less than 0.01~Hz, $P(\nu)$ is constant, which aims to
avoid the singularity at $\nu=0$. The power spectrum density for both
CMB and atmospheric noise are shown in Figure \ref{fig:PSD_CMB_ATM}. As
elucidated in Section~\ref{sub:approx of noise cov-mat}, the atmospheric
emission simulation employed in this study neglects the spatial
correlation among the scan rings, leading to a simplification of the
noise covariance. It is crucial to emphasize that the noise model we
have utilized solely takes into account temporal correlations, while
disregarding spatial correlations, as previously discussed in
section~\ref{sub:approx of noise cov-mat}.

The large-area scan utilizes a basic circular scanning strategy with a
fixed elevation of $45^\circ$. The TOD sampling frequency is fixed at
50~Hz with the circular scan enduring for a period of one minute. Under
these conditions, and with a reconstructed map resolution of
$N_{side}=128$, we obtain a total of $n_{\rm{pix}}=706$ pixels in the
pixel domain, covering approximately $f_{\rm sky}=35\%$ of the sky. To
accomplish this, the 8-hour observation is divided into 19 scan-sets,
each spanning a duration of $25$~minutes. By assuming the absence of
correlation between these scan-sets, independent solutions are derived
from each one. Eventually, a co-addition is performed on these
scan-sets, employing an inverse noise variance weight in the pixel
domain. Neglecting the variance of the Cosmic Microwave Background
(CMB), the inverse noise variance weights are determined as follows:
\begin{align}
    w_{ij, \rm c} &= \frac{[(\bm{A}^t_{j}\bm{A}_{j})^{-1}]^{-1}_{ii}}
    {\sum_j [(\bm{A}^t_{j}\bm{A}_{j})^{-1}]^{-1}_{ii}}, \;
    w_{ij, \times} = \frac{[(\bm{A}^t_j\bm{C}^{\times}\bm{A}_j)^{+}]^{-1}_{ii}}
    {\sum_{j}[(\bm{A}_j^t\bm{C}^{\times}\bm{A}_j)^{+}]^{-1}_{ii}}, \;
    w_{ij, +} = \frac{[(\bm{A}^t_j\bm{C}^{+}\bm{A}_j)^{+}]^{-1}_{ii}}
    {\sum_j [(\bm{A}^t_j\bm{C}^{+}\bm{A}_j)^{+}]^{-1}_{ii}},
\end{align}
where $i$ is the pixel index, $j$ is the scan-set index, and subscript
$ii$ denotes a matrix's $i$-th diagonal element. The final sky map is
the weighted average of the solutions from each scan-set:
\begin{align}
    \hat{x}_i = \sum_j w_{ij}\hat{x}_{ij}
\end{align}

\begin{figure}[!ht]
    \centering
    \includegraphics[width=0.32\textwidth]{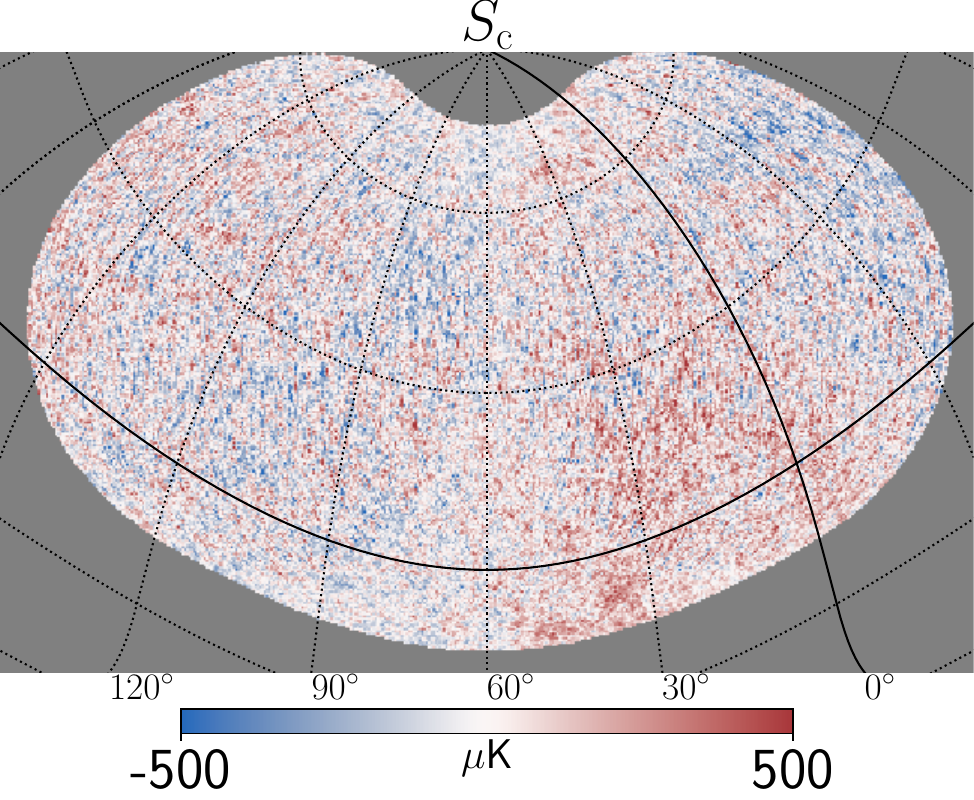}
    \includegraphics[width=0.32\textwidth]{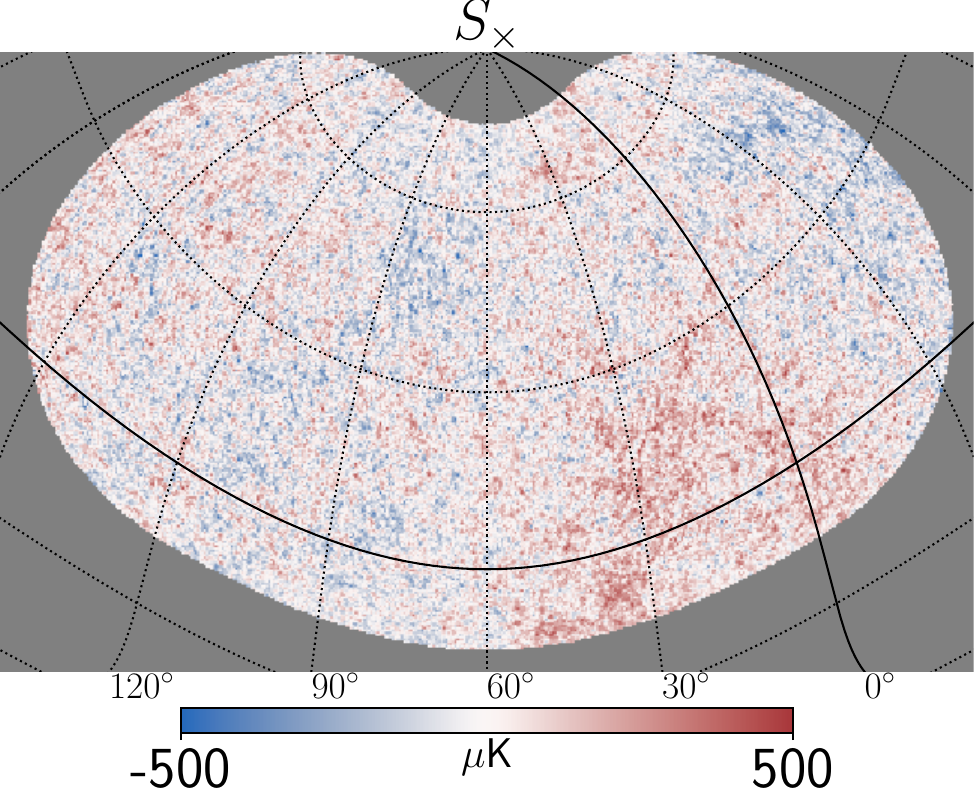}
    \includegraphics[width=0.32\textwidth]{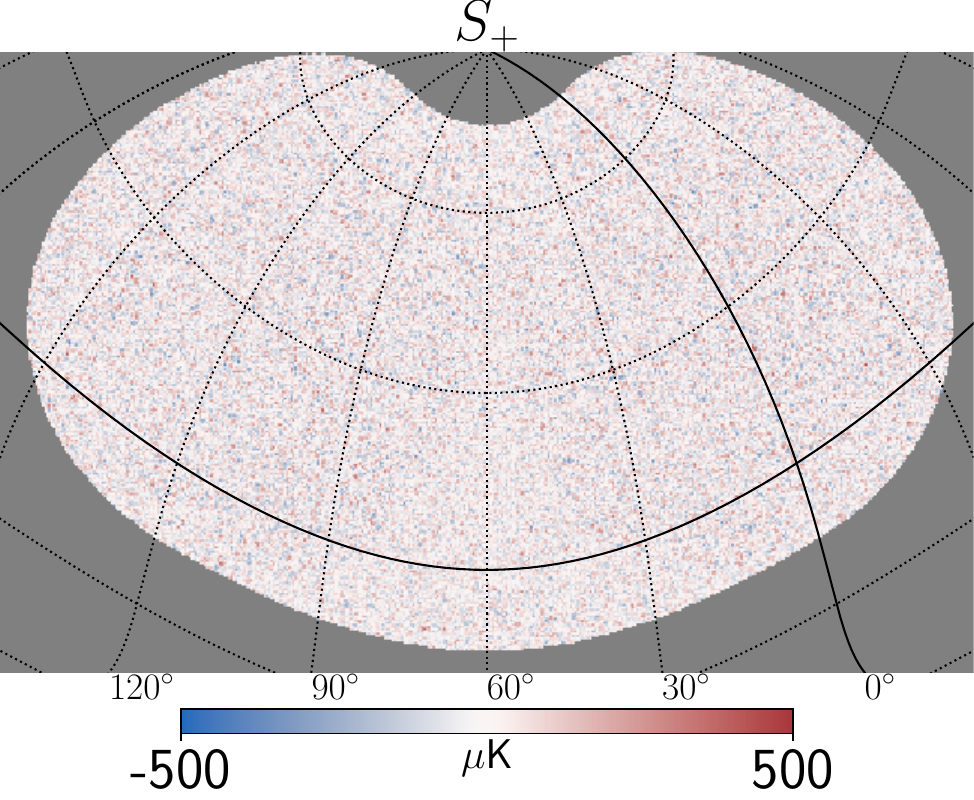}
    
    \includegraphics[width=0.32\textwidth]{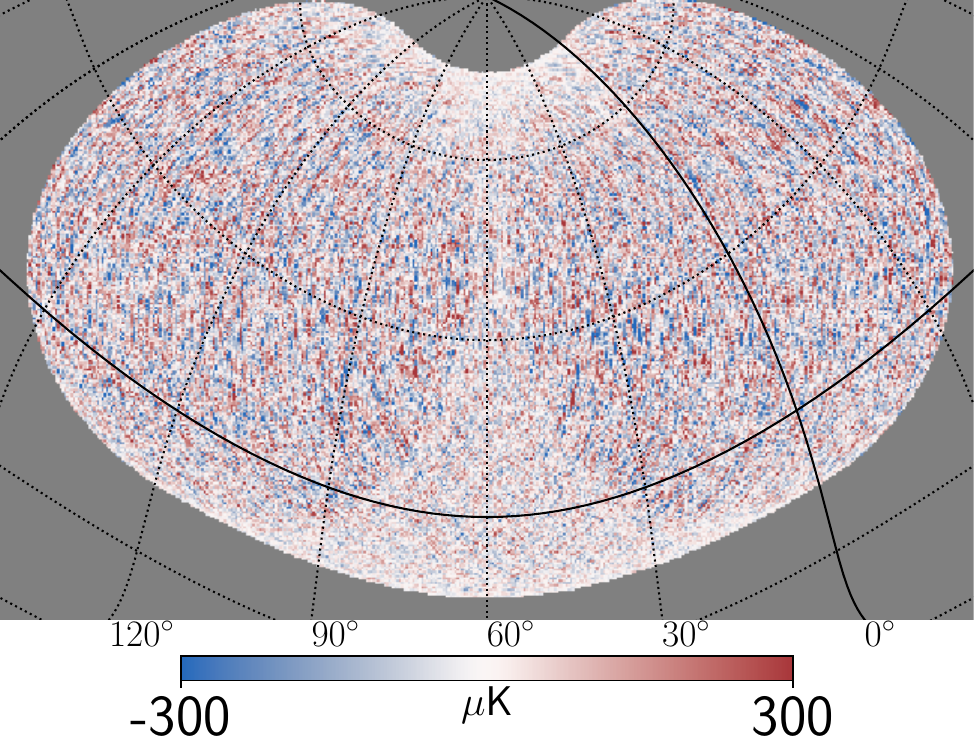}
    \includegraphics[width=0.32\textwidth]{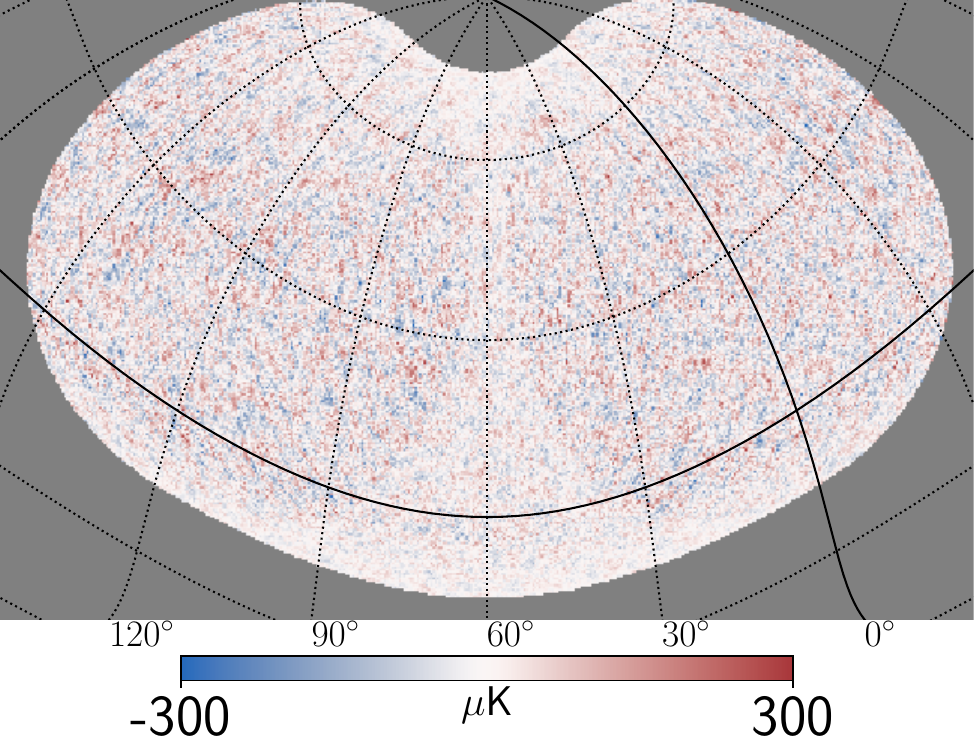}
    \includegraphics[width=0.32\textwidth]{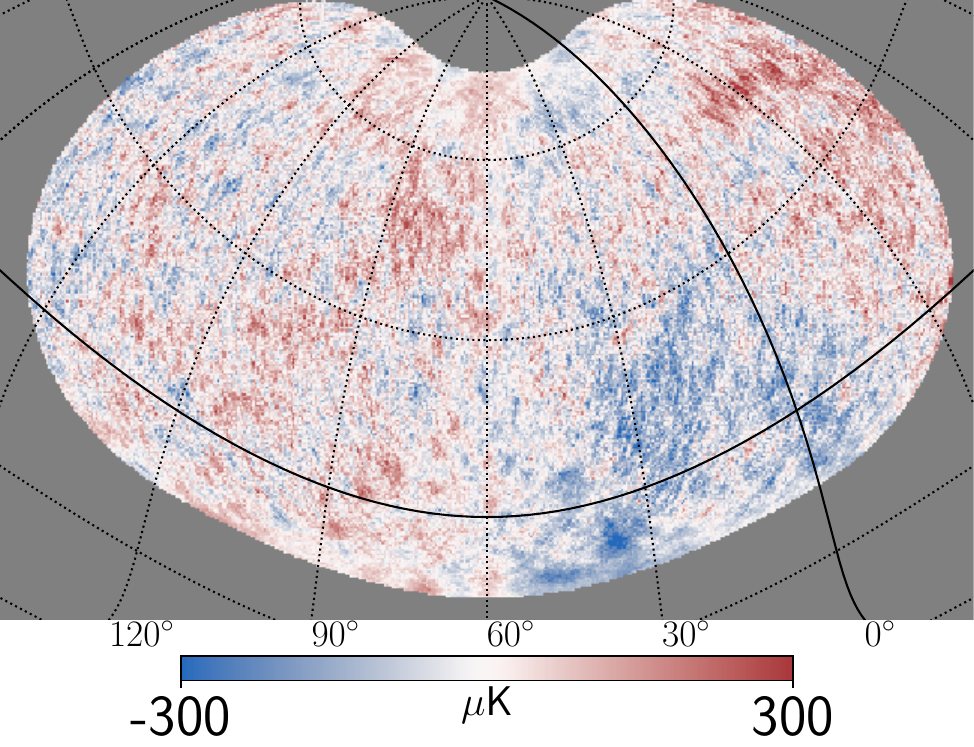}
    \caption{Upper panel: the maps reconstructed by the solutions
    $\bm{S}_{\rm c}, \bm{S}_{\times}$ and $\bm{S}_{+}$ (from left to
    right). Lower panel: the corresponding pixel domain residuals
    (output - noiseless CMB input). Obviously, the result by
    $\bm{S}_{\times}$ (middle) has the lowest overall residual
    (lower-middle). The map resolution is $N_{side}=256$.}
    \label{fig:large_maps}
\end{figure}

In the context of a singular covariance matrix $\bm{C}$, the map-level
outcomes are depicted in Figure~\ref{fig:large_maps}. The figure shows
the maps reconstructed by the solutions $\bm{S}_{\rm c},
\bm{S}_{\times}$ and $\bm{S}_{+}$, as well as their respective residuals
from left to right. It is visually apparent that the recommended
solution $\bm{S}_\times$ (lower-middle) has lower residual compared to
the other two solutions (lower-left and lower-right).

Considering the map-making procedure as a form of linear transformation
for TOD, we generated $N=400$ sets of CMB-only and noise-only
simulations, with the observed sky map for the $i$-th set given by
\begin{equation}
    \bm{x}_{i}^{\rm obs} = \bm{x}_{i}^{\rm CMB} + \bm{x}_{i}^{\rm noise}. 
\end{equation}
And the noise-debiased power spectrum for the $i$-th observation maps
$\bm{x}^{\rm obs}_{i}$ is
\begin{equation}
    \hat{C}_{\ell, i} = \hat{C}^{\rm obs}_{\ell, i} - 
    \frac{1}{N-1}\sum_{k\neq i} \hat{C}^{\rm noise}_{\ell, k}
\end{equation}
The angular power spectrum $\hat{C}_{\ell}$ is estimated from the pseudo
power spectrum $\tilde{C}_{\ell}$ by the MASTER
method~\cite{2002ApJ...567....2H} with the \texttt{NaMaster}
implementation~\cite{namaster}, where the full sky angular power
spectrum is recovered from the partial sky spectrum with an
$\ell\times\ell$ coupling matrix, determined by the sky mask with
optional apodization. In this work, the apodization is done with a
3-degree C2 configuration, and the basic angular power spectrum
computation is done using the \texttt{anafast} routine in the
\texttt{HEALPix} package~\cite{2005ApJ...622..759G}.

\begin{figure}[!ht]
    \centering
    \includegraphics[width=0.95\textwidth]{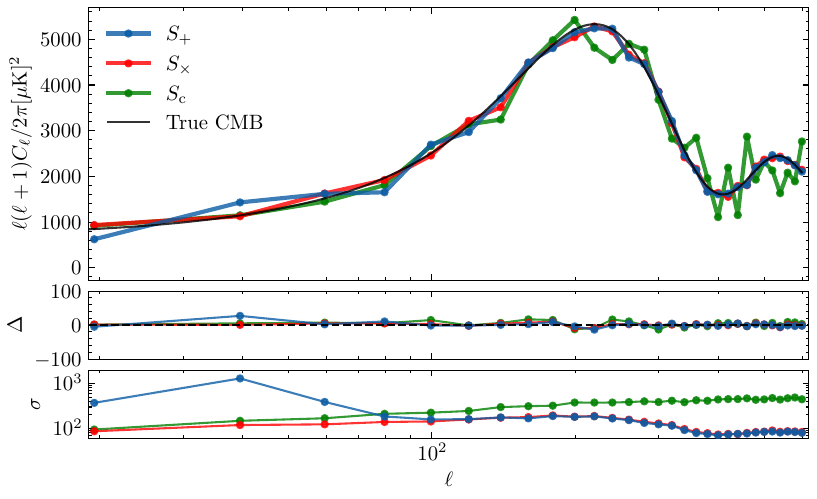}
    \caption{ The angular power spectrum results corresponding to
    Figure~\ref{fig:large_maps}, with the mean noise spectra removed and
    the suppression factors at each bin corrected. The bin width is
    $\Delta \ell = 20$ for $10\leq\ell<610$. Top panel: the corrected
    spectra of one realization as example. Middle panel: difference
    between the expected CMB angular power spectrum and each solutions'
    mean spectrum of 400 simulations. Bottom panel: the corresponding
    error bar amplitudes at each $\bm{\ell}$. }
    \label{fig:large_powers}
\end{figure}

The average and variance of the angular power spectra with bin width
$\Delta\ell=20$ of the solutions $\bm{S}_c, \bm{S}_\times$ and
$\bm{S}_+$ are depicted in Figure~\ref{fig:large_powers} using green,
red, and blue solid lines, respectively; which shows that, after
eliminating the average noise spectrum and correcting the suppression
effect arising from filtering (pertaining exclusively to the $\bm{S}_+$
solution) and masking (pertaining to all solutions), unbiased angular
power spectra can be obtained from all solutions. Notably, the
$\bm{S}_\times$ solution displays significantly reduced error bar
amplitudes in comparison to both the $\bm{S}_{+}$ solution (for
$\ell<80$) and the $\bm{S}_{\rm c}$ solution (for $\ell>30$). Therefore,
the $\bm{S}_\times$ solution outperforms the other two solutions across
the majority of $\ell$-ranges.

\subsection{The performance tests: for a small sky region}\label{subsec:small}

In this validation procedure, we focus on a relatively small sky region,
covering approximately $f_{\rm sky} = 5\%$. We also used observations
with a duration of 8 hours, but they were divided into 16 scan-sets,
each lasting 30 minutes. Similarly, we ignore the correlation between
scan-sets.

In actual observational experiments, the scanning strategy is not
circular. Instead, the telescope scans back-and-forth at a fixed
elevation angle. In this case, we pay more attention to the spatial
correlation of atmospheric emission.

Assuming the atmosphere is a vast 2D plane and the power spectrum of
the emission intensity satisfies $P(\bm{k}) =
\left<f(\bm{k})f^*(\bm{k}') \right>\propto |\bm{k}|^{-3}\delta(\bm{k} -
\bm{k}')$, where $f(\bm{k})$ is the 2D Fourier transformation of the
atmospheric emission intensity, the covariance in real domain between
any two points can be derived trivially:
\begin{align}
    {\rm Corr}(\delta r) = \left<T(\bm{r}) T(\bm{r} + 
    \delta \bm{r})\right> \propto \int_{0}^\infty {\rm d} k k^{-2} J_0(k\delta r),
\end{align}
where $J_0$ is the first kind Bessel function. The covariance can be
obtained through numerical computation. To avoid integral divergence,
we also set a cutoff frequency, below which the power spectrum of
atmospheric emission is constant. This is similar to what is shown in
Figure \ref{fig:PSD_CMB_ATM}.

The physical distance between two sample points can be easily
calculated from scan strategy data. Taking into account the influence
of wind on the atmosphere, the covariance on TOD domain can be written
as:
\begin{align}
    C_{ij} = {\rm Corr}(\big\lvert \delta\bm{r}_{ij} + \bm{v}_w\delta
    t_{ij} \big\rvert), 
\end{align}
where $\delta\bm{r}_{ij}, \delta t_{ij}$ denote the physical distance
and time difference between two samples, respectively. $\bm{v}_w$
denotes the wind speed. For the sake of simplicity, it is treated as a
constant in the simulation.

Any linear filter $\bm{M}$ operating in the TOD domain will also distort
the covariance matrix. The relationship between the filtered and
unfiltered covariance is simply $\tilde{\bm{C}} =
\bm{M}\bm{C}\bm{M}^t$. Here, we also employ a high-pass filter with the
same parameters as in the previous section. The only difference is that
the filter operates on each half-scan independently, which means the
filter matrix $\bm{M}$ is a block diagonal matrix.

\begin{figure}[!ht]
    \centering
    \includegraphics[width=0.32\textwidth]{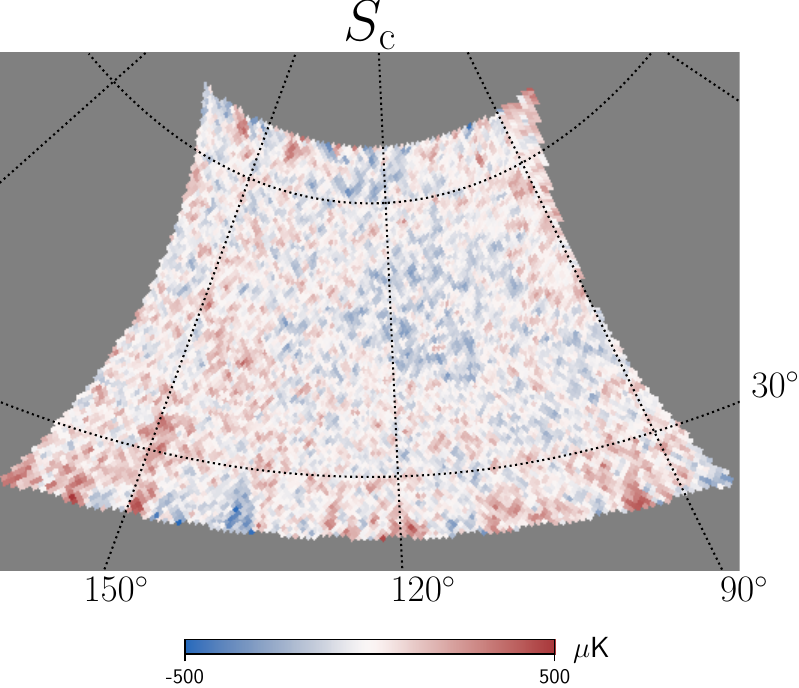}
    \includegraphics[width=0.32\textwidth]{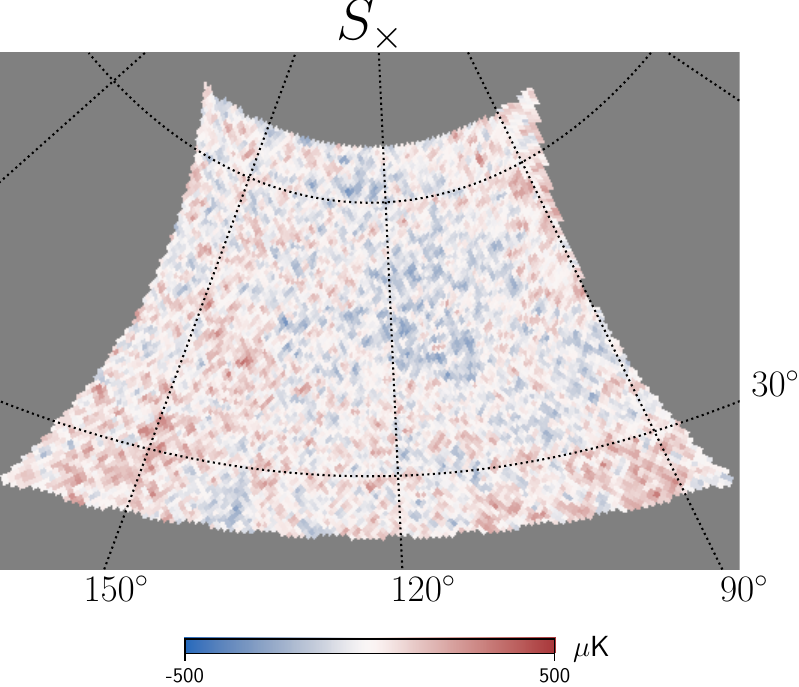}
    \includegraphics[width=0.32\textwidth]{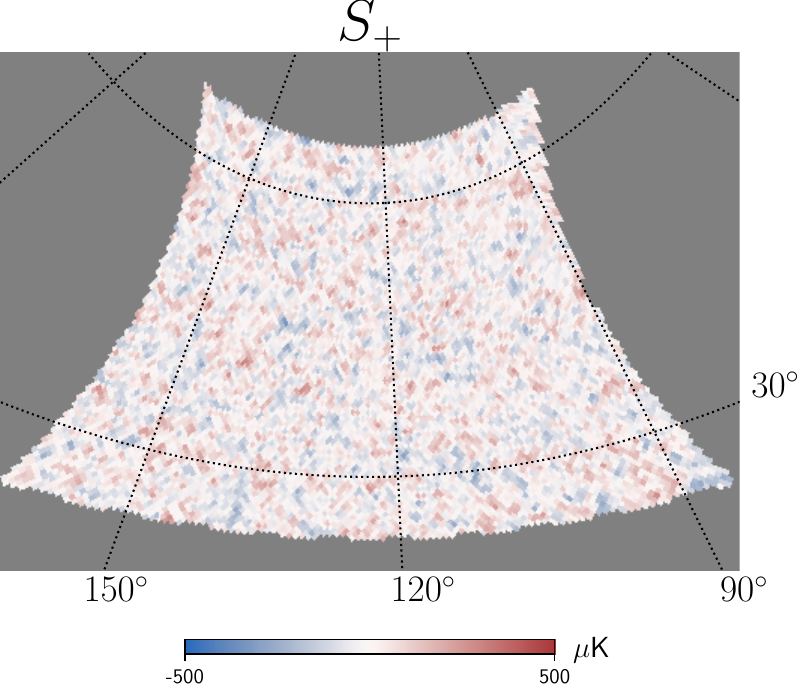}
    
    \includegraphics[width=0.32\textwidth]{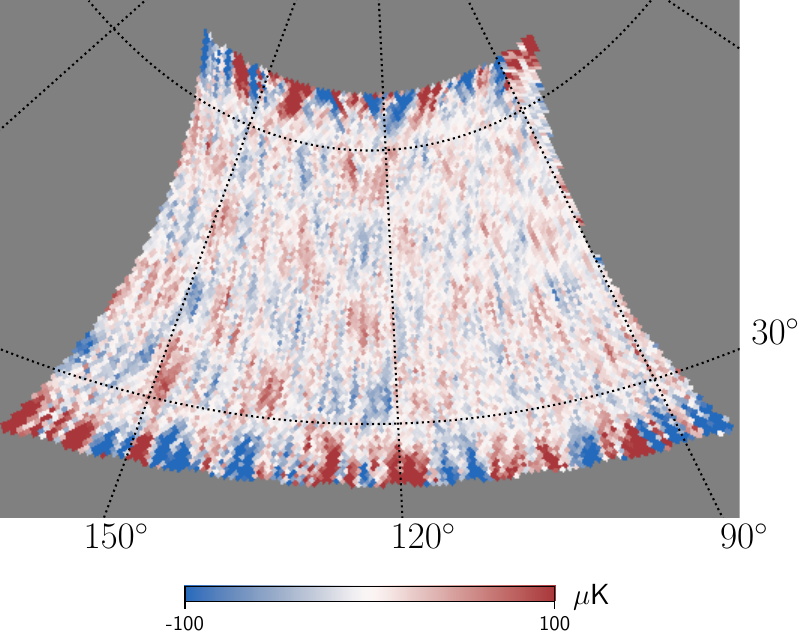}
    \includegraphics[width=0.32\textwidth]{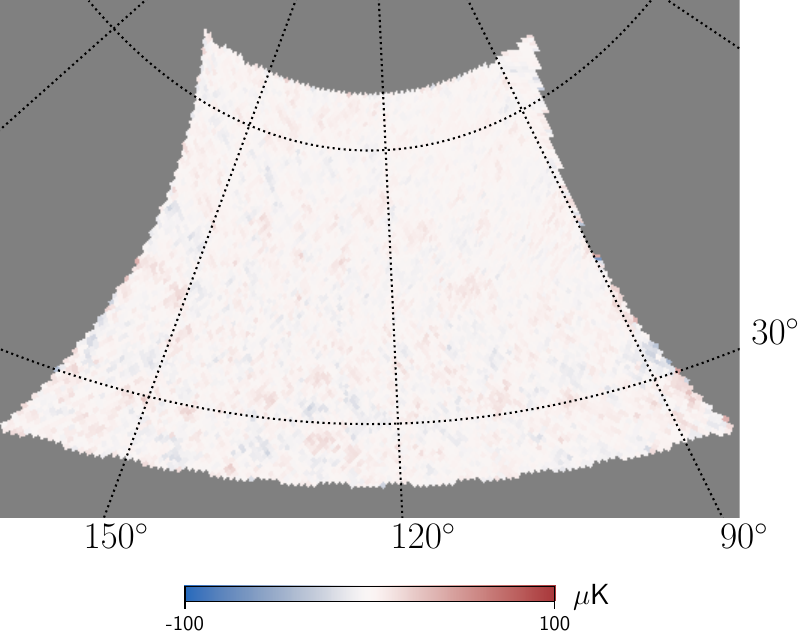}
    \includegraphics[width=0.32\textwidth]{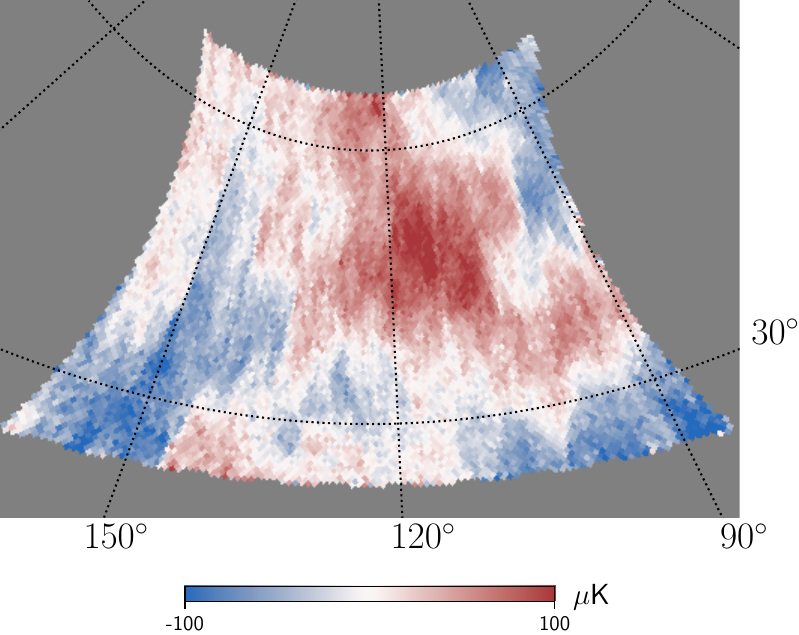}
    \caption{The reconstructed map and the corresponding pixel domain
    residuals. Figure conventions are similar to those of Figure
    \ref{fig:large_maps}. Obviously, the result by $\bm{S}_{\times}$
    (middle) has the lowest overall residual (lower-middle). The map
    resolution is $N_{side}=128$.}
    \label{fig:small_maps}
\end{figure}

\begin{figure}[!ht]
    \centering
    \includegraphics[width=0.95\textwidth]{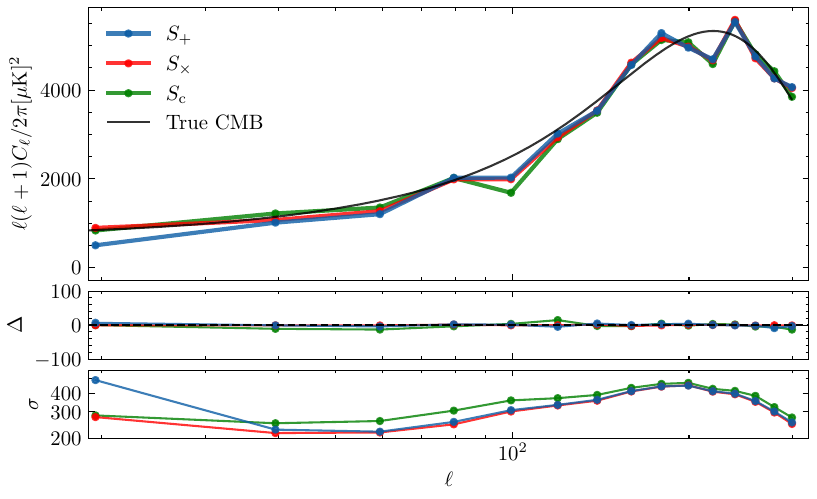}
    \caption{ The angular power spectrum results corresponding to
    Figure~\ref{fig:small_maps}, with the mean noise spectra removed and
    the suppression factors at each $\ell$ corrected. The bin width is
    $\Delta \ell = 20$ for $10\leq\ell<310$. Top panel: the corrected
    spectra of one realization as example. Middle panel: difference
    between the expected CMB angular power spectrum and each solutions'
    mean spectrum of 400 simulations. Bottom panel: the corresponding
    error bar amplitudes at each $\bm{\ell}$. }
    \label{fig:small_powers}
\end{figure}

The subsequent map-making processing for the case of a small sky region
is similar to that mentioned in the previous section.
Figure~\ref{fig:small_maps} displays the solutions $\bm{S}_c,
\bm{S}_\times$, and $\bm{S}_+$ with their residuals from left to right
at the map-level. Similar to Figure~\ref{fig:large_maps} and even more
evident, the $\bm{S}_\times$ solution has the lowest residual
(lower-middle), while the residual of $\bm{S}_c$ (lower-left) is
dominated by instrumental noise and the residual of $\bm{S}_+$
(lower-right) is dominated by large-scale CMB loss due to filtering. The
angular power spectra results corresponding to
Figure~\ref{fig:small_maps} are illustrated in
Figure~\ref{fig:small_powers}. It is notable that the solution
$\bm{S}_\times$ achieves the best error bar across almost all angular
scales.

\section{Summary and discussions}
\label{sec:disscuss}

In this paper, we show how to solve for the sky map in an optimal way
with a full consideration of the singularity problems. In summary, to
get the true optimal solution, the singular value decomposition should
be employed along with a detailed investigation of the effective noise
covariance matrix, which is the most complex part of the solution. Once
this is done as indicated in the appendix, the rest of the solution can
be done with a regular singular-value decomposition to discover the
general true optimal solution. This approach also tells us, that the
general true optimal solution depends on the situation of singularity,
and cannot be achieve with just one simple equation.

Considering the complexity of the true optimal solution, we have also
designed a simplified sub-optimal solution that can be given in just one
simple equation, which is only slightly different from the classical
pseudo inverse method and is hence much easier to implement. The side
effect of this sub-optimal solution is strictly limited to the singular
part, which is usually negligible. Moreover, this sub-optimal solution
also helps to greatly suppress the unwanted noise amplification effect,
as discussed in section~\ref{sub:regarding the main scan}. Simulations
with the AliCPT observation strategies confirm that this sub-optimal
solution is capable of producing significantly better sky maps as well
as angular power spectra than the naive and classical pseudo inverse
solutions, especially at the large scales.

As a possible augment of the method in this work, one may even solve for
the best estimation of the CMB angular power spectrum directly from the
TOD, which means to re-write the main equation into
\begin{align}
    \bm{d} = \bm{A}\bm{M}\bm{Y}\bm{a}+\bm{n},
\end{align}
where $\bm{M}$ is a mask corresponding to the observation region,
$\bm{Y}$ is a matrix of the spherical harmonics, and $\bm{a}$ is the
spherical harmonic coefficients, which satisfy $\langle \bm{a}\bm{a}^t
\rangle = \bm{C}_\ell$ (diagonal), i.e., the covariance matrix of
$\bm{a}$ is the CMB angular power spectrum. Then we can either solve for
the contribution of $\bm{d}$ to $\bm{a}$ by the method described in this
work, or even use an quadratic maximum likelihood (QML) method to solve
for the best $\bm{C}_\ell$, because the covariance matrix of $d$ is
\begin{align}
    \langle \bm{d} \bm{d}^t \rangle = \bm{A}\bm{M}\bm{Y}\bm{C}_\ell
    \bm{Y}^H\bm{M}\bm{A}^t + \langle \bm{n}\bm{n}^t \rangle,
\end{align}
whose partial derivative to $C_\ell$ is easy to obtain, and hence
enables the QML calculations. Ideally, this can give the best estimation
of $C_\ell$ even without map-making; however, because we need to work in
the pixel domain for the foreground removal and delensing, such a direct
solution of $C_\ell$ is not a good choice unless those pixel domain
operations can be successfully transported to the time domain, which is
an exciting idea but does not seems to be feasible in the near future.
However, we still aim to test a low-resolution version of this idea as
an interesting follow-up of this work.

The concept of the near-optimal solution can readily assist in the
solution of singularity issues in other approaches, such as the maximum
likelihood solution, so long as the method calls for the inversion of a
singular covariance matrix. Despite the difficulty of obtaining the true
optimal solution of the singularity problem, it is easy to incorporate
the recommended sub-optimal solution in other approaches: when one needs
to invert a singular noise covariance matrix, the inversion should done
with a modified pseudo-inverse as explained in
appendix~\ref{subsub:modified pseudo inverse}, and other matrix
inversions can be done with the regular inverse or pseudo-inverse (if
necessary). By this simple operation, the benefits of the recommended
solution is fully realized, even in another approach.

\Ack{

This work is supported by the National Key R\&D Program of China No.
2021YFC2203100 and No. 2021YFC2203104, the Anhui Provincial Natural
Science Foundation 2308085MA30 and the Anhui project Z010118169. We 
thank the anonymous referee for the useful comments and James Creswell 
for helping improve this work.

}

\appendix

\section{The mathematical details}\label{app:math details}

\subsection{How to deal with the singularity of the noise covariance matrix}
\label{sub:deal with singular cov}

To deal with a singular noise covariance matrix might be the most
complicated part, but in most cases, the solution can be eventually
simplified. We shall first introduce the general method that is somewhat
complex, and then discuss simplifications to the method.

Because the covariance matrix $\bm{C} = \< \bm{n}\bm{n}^t \>$ does
converge, it can always be decomposed into $\bm{C} = \bm{P}
\bm{\lambda}^2 \bm{P}^t$, where $\bm{P}$ is unitary and $\bm{\lambda}$
is diagonal (allowing zero diagonal elements). Therefore, it is always
possible to convert $\bm{n}$ into a standard white noise $\bm{n}_0$ that
follows the standard normal distribution and satisfy $\<\bm{n}_0
\bm{n}_0^t\>=\bm{I}$. The connection is given by $\bm{n} =
\bm{P}\bm{\lambda}\bm{n}_0$, which automatically gives $\bm{C} = \<
\bm{n}\bm{n}^t \> = \bm{P}
\bm{\lambda}^2 \bm{P}^t$; thus, we get
\begin{align}\label{equ:a1}
\bm{d} &= \bm{A} \bm{x} + \bm{P} \bm{\lambda} \bm{n}_0 \\ \nonumber
\bm{P}^t \bm{d} &= \bm{P}^t\bm{A} \bm{x} + \bm{\lambda} \bm{n}_0.
\end{align}
Because $\bm{\lambda}$ may have zero diagonal elements, we divide
$\bm{\lambda}$ and $\bm{n}_0$ into blocks as
\begin{align}
\bm{\lambda} = \begin{pmatrix}
\bm{\lambda}_1 & \bm{0} \\
\bm{0} & \bm{0} \\
\end{pmatrix}, \; \bm{n}_0 = \begin{pmatrix}
\bm{n}_1 \\ \bm{0}
\end{pmatrix},
\end{align}
where $\bm{\lambda}_1$ is diagonal and invertible, and the number of
zero rows is $\ell$. Correspondingly, we divide $\bm{P}^t$ into two
blocks with $n_{\rm tod}-\ell$ and $\ell$ rows, respectively:
\begin{align}
\bm{P}^t = \begin{pmatrix}
\bm{P}^t_1 \\ \bm{P}^t_2
\end{pmatrix} \;\longrightarrow\; \bm{P}^t\bm{d} =
\begin{pmatrix} \bm{P}^t_1\bm{d} \\ \bm{P}^t_2\bm{d} \end{pmatrix} = 
\begin{pmatrix} \bm{d}_1 \\ \bm{d}_2 \end{pmatrix}.
\end{align}
Thus, eq.~(\ref{equ:a1}) becomes
\begin{align}\label{equ:adfaklj}
\begin{pmatrix} \bm{d}_1 \\ \bm{d}_2 \end{pmatrix} =
\begin{pmatrix} \bm{P}^t_1\bm{A}\bm{x} \\ \bm{P}^t_2\bm{A}\bm{x}
\end{pmatrix} + 
\begin{pmatrix} \bm{\lambda}_1\bm{n}_1 \\ \bm{0}
\end{pmatrix}.
\end{align}
Evidently, this equation should first be solved for the lower part
because it effectively contains no noise:
\begin{align}\label{equ:0277623}
\bm{d}_2 = (\bm{P}^t_2\bm{A})\bm{x}.
\end{align}
However, $\bm{P}^t_2\bm{A}$ has $\ell$ rows and $n_{\rm{pix}}$ columns,
and $\ell$ is almost always much smaller than $n_{\rm{pix}}$, so the
number of unknowns is bigger than the number of conditions. Therefore,
the equation cannot be solved directly. Here we first take the singular
value decomposition of $\bm{P}^t_2\bm{A}$ as
\begin{align}\label{equ:092384asdf2}
\bm{P}^t_2\bm{A} = \bm{F}\bm{\Lambda}\bm{G}^t,
\end{align}
where $\bm{F}$ is $\ell \times \ell$ and unitary; $\bm{G}$ is
$n_{\rm{pix}} \times n_{\rm{pix}}$ and unitary; and $\bm{\Lambda}$ has
$\ell$ rows and $n_{\rm{pix}}$ columns. However, most elements of
$\bm{\Lambda}$ are zero, except for an $\ell \times \ell$ square block
$\bm{\Lambda}_1$ on its top-left containing the singular values of
$\bm{P}^t_2\bm{A}$.

Let $\bm{x}_1 = \bm{G}^t\bm{x}$ be a column vector, whose first $\ell$
elements are denoted as $\bm{x}_{1,\rm{upper}}$, and the rest,
$\bm{x}_{1,\rm{lower}}$. We write $\bm{P}^t_2\bm{A}\bm{x}$ explicitly
as:
\begin{align}
\bm{d}_2=\bm{P}^t_2\bm{A}\bm{x} =\bm{F} \begin{pmatrix}
\bm{\Lambda}_1 & \bm{0}
\end{pmatrix} \bm{G}^t \bm{x} = \bm{F} \begin{pmatrix}
\bm{\Lambda}_1 & \bm{0}
\end{pmatrix} \begin{pmatrix}
\bm{G}_1^t\bm{x} \\ \bm{G}_2^t\bm{x}
\end{pmatrix} = \bm{F} \begin{pmatrix}
\bm{\Lambda}_1 & \bm{0}
\end{pmatrix} \begin{pmatrix}
\bm{x}_{1,\rm{upper}} \\ \bm{x}_{1,\rm{lower}}
\end{pmatrix} = \bm{F} \bm{\Lambda}_1 \bm{x}_{1,\rm{upper}}.
\end{align}
Thus, eq.~(\ref{equ:0277623}) becomes
\begin{align}\label{equ:092384alk}
\bm{d}_2 = \bm{F} \bm{\Lambda}_1
\bm{x}_{1,\rm{upper}} \;\;\longrightarrow\;\;
\bm{x}_{1,\rm{upper}} = \bm{\Lambda}_1^{-1}\bm{F}^t\bm{d}_2,
\end{align}
which is a precise solution (noiseless). Now we can return to
eq.~(\ref{equ:adfaklj}) to deal with its upper half: $\bm{d}_1 =
\bm{P}_1^t\bm{A}\bm{x} + \bm{\lambda}_1\bm{n}_1$. Substituting $\bm{x} =
\bm{G}\bm{x}_1$, we get
\begin{align}\label{equ:98a798a7}
\bm{d}_1 = (\bm{P}_1^t\bm{A}\bm{G})\bm{x}_1 + \bm{\lambda}_1\bm{n}_1 =
\bm{A}_1\begin{pmatrix}
\bm{x}_{1,\rm{upper}} \\ \bm{x}_{1,\rm{lower}}
\end{pmatrix} + \bm{\lambda}_1\bm{n}_1,
\end{align}
Because $\bm{x}_{1, upper}$ is solved precisely, we divide $\bm{A}_1$
into blocks with $\ell$ and $n_{\rm{pix}}-\ell$ columns, respectively:
\begin{align}
\bm{A}_1 = \begin{pmatrix}
    \bm{A}_{1, \rm{left}} & \bm{A}_{1, \rm{right}}
\end{pmatrix},
\end{align}
then we have
\begin{align}\label{equ:83akjiewry}
\bm{d}_1 = \begin{pmatrix}
\bm{A}_{1, \rm{left}} & \bm{A}_{1, \rm{right}}
\end{pmatrix}\begin{pmatrix}
\bm{x}_{1,\rm{upper}} \\
\bm{x}_{1,\rm{lower}}
\end{pmatrix} + \bm{\lambda}_1\bm{n}_1 =
\bm{A}_{1, \rm{left}}\bm{x}_{1,\rm{upper}} + 
\bm{A}_{1, \rm{right}}\bm{x}_{1,\rm{lower}} +
\bm{\lambda}_1\bm{n}_1. 
\end{align}
Because $\bm{x}_{1,\rm{upper}}$ is already known, we get
\begin{align}
\bm{d}_1 - \bm{A}_{1, \rm{left}}\bm{x}_{1,\rm{upper}} = \bm{d}_1' 
= \bm{A}_{1,
\rm{right}}\bm{x}_{1,\rm{lower}} + \bm{\lambda}_1\bm{n}_1.
\end{align}
The noise covariance matrix in the above equation is apparently
non-singular. Therefore, the problem of singular covariance matrix is
solved, and in the next step, we can start with the above equation that
has no covariance matrix singularity. Once $\bm{x}_{1, \rm{lower}}$ is
solved, we immediately get $\bm{x}=\bm{G}\bm{x}_1$.

\subsection{How to safely reduce the size of the linear system}
\label{sub:reducing the size}

In the above section, we have explained how to safely convert the
problem with a singular covariance matrix to one with invertible
covariance matrix; therefore, below we will assume the noise covariance
matrix is invertible, and the noise is already converted to standard
white noise like:
\begin{align}\label{equ:asdfi9223}
(\bm{\lambda}^\times \bm{P}^t \bm{d}) &= (\bm{\lambda}^\times \bm{P}^t\bm{A})
\bm{x} + \bm{n}_0.
\end{align}
For convenience, we rewrite the above equation and \emph{renew} the
symbols $\bm{A}$ and $\bm{d}$, so as to continue with the equation
below:
\begin{align}\label{equ:asdfi9222131asdf}
\bm{d} &= \bm{A}\bm{x} + \bm{n}_0.
\end{align} 

To obtain the solution of eq.~(\ref{equ:asdfi9222131asdf}), we first
write the singular value decomposition (SVD) of $\bm{A}$ as
\begin{align}\label{equ:svd A}
\bm{A} = \bm{U} \bm{D} \bm{V}^t,
\end{align}
where $\bm{U}$ has the same shape as $\bm{A}$ and $\bm{U}^t\bm{U}$ is an
$n_{\rm{pix}}\times n_{\rm{pix}}$ identity matrix, $\bm{D}$ is an
$n_{\rm{pix}}\times n_{\rm{pix}}$ diagonal matrix containing the
singular values of $\bm{A}$, and $\bm{V}$ is $n_{\rm{pix}}\times
n_{\rm{pix}}$ and unitary. Substituting the decomposition into
eq.~(\ref{equ:asdfi9222131asdf}) shows
\begin{align}\label{equ:point of simp}
\bm{U}^t\bm{d} &= \bm{D}\bm{V}^t\bm{x} + \bm{U}^t\bm{n}_0.
\end{align}
This operation targets at reducing the size of the problem because, in
the above equation, the sizes of the column vectors $\bm{U}^t\bm{d}$,
$\bm{V}^t\bm{x}$ and $\bm{U}^t\bm{n}$ are all $n_{\rm{pix}}$, which is
much smaller than $n_{\rm tod}$. Meanwhile, because $\bm{U}$ comes from
the singular value decomposition of $\bm{A}$, the above step caused no
loss of the sky map information from $\bm{A}\bm{x}$ to
$\bm{D}\bm{V}^t\bm{x}$.

However, there is another concern: Because $\bm{U}$ is not an unitary
matrix (non-square), when we left-multiply $\bm{U^t}$ to both sides of
the equation, it does change the noise information. In order to explain
this issue, we introduce another equivalent convention of the singular
value decomposition that shows
\begin{align}\label{equ:svd A 1}
\bm{A} = \bm{U}_1 \bm{D}_1 \bm{V}^t_1,
\end{align}
where $\bm{U}_1$ is $n_{\rm tod} \times n_{\rm tod}$ and unitary;
$\bm{D}_1$ has $n_{\rm tod}$ rows and $n_{\rm{pix}}$ columns, but most
of its elements are zero, except for the $n_{\rm{pix}}\times
n_{\rm{pix}}$ square block on top of it, which is diagonal and contains
the singular values of $\bm{A}$; and $\bm{V}_1$ is still $n_{\rm{pix}}
\times n_{\rm{pix}}$ and unitary. Compared with eq.~(\ref{equ:svd A}),
this form appends normalized and linearly independent columns to the
right of $\bm{U}$ to make it unitary, and zeros to the bottom of
$\bm{D}$ to match the rule of matrix multiplication, but does not change
the matrix multiplication. Thus, the two forms of singular value
decomposition are equivalent.

The advantage of eq.~(\ref{equ:svd A 1}) is that $\bm{U}_1$ is unitary;
thus it is always safe to left-multiply $\bm{U}_1^t$ to both sides of
eq.~(\ref{equ:asdfi9222131asdf}) to give
\begin{align}\label{equ: point of simp 1}
\bm{U}_1^t\bm{d} = \bm{D}_1 \bm{V}_1^t \bm{x} + \bm{U}_1^t \bm{n}_0.
\end{align}
The above equation contains many more rows than eq.~(\ref{equ:point of
simp}); however, it is easy to observe the following facts:
\begin{enumerate}
\item \label{itm:asdf21q32} The top $n_{\rm{pix}}$ lines are identical
between eq.~(\ref{equ:point of simp} \& \ref{equ: point of simp 1}).

\item \label{itm:af2234121} The remaining lines in eq.~(\ref{equ: point of
simp 1}) contain only noise and no information from the sky map.

\item \label{itm:adfa5asdf} The covariance matrix of $\bm{U}_1^t
\bm{n}_0$ is $\<(\bm{U}_1^t
\bm{n}_0)(\bm{U}_1^t \bm{n}_0)^t\>=\bm{I}$, i.e., $\bm{U}_1^t \bm{n}_0$
is another realization of standard white noise.
\end{enumerate}
Items~\ref{itm:asdf21q32},~\ref{itm:af2234121} ensure that the sky map
information should be solved from the first $n_{\rm{pix}}$ equations,
which is exactly eq.~(\ref{equ:point of simp}). Item~\ref{itm:adfa5asdf}
ensures that the noise terms in the remaining $n_{\rm tod}-n_{\rm{pix}}$
equations are uncorrelated with the noise terms in the first
$n_{\rm{pix}}$ equations; thus, they cannot be used to cancel any noise
component in the first $n_{\rm pix}$ rows, i.e., they cannot help to
improve the signal-to-noise ratio (SNR). Therefore, the optimal solution
given by eq.~(\ref{equ:point of simp}, \ref{equ: point of simp 1}) are
identical.

\subsection{How to deal with the main matrix's singularity}
\label{sub: main matrix singularity}

Because the main matrix $\bm{A}$ is non-square, its singularity is
better represented by the singularity of $\bm{A}^t\bm{A} = \bm{V}
\bm{D}^2 \bm{V}^t$. When the latter square matrix is singular, there
could be zero diagonal elements in $\bm{D}$, like:
\begin{align}\label{equ:asdfadsfasdf23}
\bm{D} = \begin{pmatrix}
\bm{D}_1 & \bm{0} \\
\bm{0} & \bm{0}
\end{pmatrix},
\end{align}
where $\bm{D}_1$ is an invertible diagonal matrix, the number of zero
rows is $k$, and all $\bm{0}$s are zero matrices. We then divide the
column vectors in eq.~(\ref{equ:point of simp}) as follows:
\begin{align}\label{equ:block def}
\bm{U}^t\bm{d} = \begin{pmatrix}
\bm{d}_1 \\ \bm{*} 
\end{pmatrix},
\bm{V}^t\bm{x} = \begin{pmatrix}
\bm{x}_1 \\ \bm{*} 
\end{pmatrix},
\bm{U}^t\bm{n}_0 = \begin{pmatrix}
\bm{n}_1 \\ \bm{*} 
\end{pmatrix},
\end{align}
where ``$\bm{*}$'' corresponds to the zero rows of $\bm{D}$, and
eq.~(\ref{equ:point of simp}) becomes
\begin{align} \begin{pmatrix}
\bm{d}_1 \\ \bm{*} 
\end{pmatrix} =  \begin{pmatrix}
\bm{D}_1 & \bm{0} \\ \bm{0} & \bm{0}
\end{pmatrix}
\begin{pmatrix}
\bm{x}_1 \\ \bm{*} 
\end{pmatrix} +  \begin{pmatrix}
\bm{n}_1 \\ \bm{*} 
\end{pmatrix} =  \begin{pmatrix}
\bm{D}_1\bm{x}_1 \\ \bm{0} 
\end{pmatrix} +  \begin{pmatrix}
\bm{n}_1 \\ \bm{*} 
\end{pmatrix}. 
\end{align}
As explained in section~\ref{sub:reducing the size}, because
$\bm{U}^t\bm{n}_0$ is another realization of standard white noise,  the
``$\bm{*}$'' rows can be safely removed from the solution; thus the
above equation is equivalent to
\begin{align}
\bm{d}_1 = \bm{D}_1 \bm{x}_1 + \bm{n}_1.
\end{align}

Because $\bm{D}_1$ is square, diagonal and invertible, it is easy to
prove that the optimal solution of the above equation is simply
\begin{align}\label{equ:step2 solu}
\widetilde{\bm{x}}_1 = \bm{D}_1^{-1} \bm{d}_1.
\end{align}
However, note that this solution is in a specific space associated with
the unitary matrix $\bm{V}$, and we eventually want to solve the sky map
$\bm{x}$ in the ordinary pixel domain.

From eq.~(\ref{equ:step2 solu}) we see that
\begin{align}
\widetilde{\bm{x}}_1 = \bm{D}_1^{-1} \bm{d}_1 = \bm{x}_1 + \bm{D}_1^{-1}
\bm{n}_1.
\end{align}
With a similar block scheme as eq.~(\ref{equ:block def}), we define
\begin{align}
\bm{V}^t = \begin{pmatrix}
\bm{V}_1^t \\ \bm{*} 
\end{pmatrix}, 
\bm{U}^t = \begin{pmatrix}
\bm{U}_1^t \\ \bm{*} 
\end{pmatrix},
\end{align}
where $\bm{V}_1^t$ is a matrix consisting of the first $n_{\rm{pix}}-k$
rows of $\bm{V}^t$, $\bm{U}_1^t$ is a matrix consisting of the first
$n_{\rm{pix}}-k$ rows of $\bm{U}^t$, and $k$ is the number of zero rows
in eq.~(\ref{equ:asdfadsfasdf23}); so we have
\begin{align}
\bm{x}_1 = \bm{V}_1^t \bm{x}, \;
\bm{n}_1 = \bm{U}_1^t \bm{n},
\end{align}
and we get the following linear equations that consist of $n-k$ rows to
solve $\bm{x}$ from.
\begin{align}\label{equ:ss3}
\widetilde{\bm{x}}_1 = \bm{V}_1^t \bm{x} + \bm{D}_1^{-1}
\bm{U}_1^t \bm{n}.
\end{align} 

The goal is to estimate $\bm{x}$ from the above equation, and the main
difficulty is that $\bm{V}_1^t$ has $n_{\rm{pix}}-k$ rows but
$n_{\rm{pix}}$ columns, which is insufficient to solve $\bm{x}$ fully.
However, because $\widetilde{\bm{x}}_1$, $\bm{V}_1^t \bm{x}$ and
$\bm{D}_1^{-1} \bm{U}_1^t \bm{n}$ are all column vectors of size
$n_{\rm{pix}}-k$, there is no problem to add $k$ zero elements to the
bottom of each of them to expand their sizes from $n_{\rm{pix}}-k$ to
$n_{\rm{pix}}$. Then we can rewrite the equation as
\begin{align}\label{equ:ss4}
\widetilde{\bm{x}}_2 = \bm{V}^t \bm{x} + \bm{n}_2,
\end{align}
where
\begin{align}
\widetilde{\bm{x}}_2 = 
\begin{pmatrix}
    \widetilde{\bm{x}}_1 \\ \bm{0}
\end{pmatrix},
\bm{n}_2 = \begin{pmatrix}
    \bm{D}_1^{-1} \bm{U}_1^t \bm{n} \\ \bm{0}
\end{pmatrix},
\end{align}
and we require a special constraint on $\bm{x}$ that its inner products
with the last $k$ rows of $\bm{V}^t$ (which is an $n_{\rm{pix}}\times
n_{\rm{pix}}$ unitary matrix) are zero, which means the corresponding
modes are supposed to be missing in $\bm{x}$.

Because $\bm{V}^t$ is $n_{\rm{pix}}\times n_{\rm{pix}}$ and unitary, it
is easy to prove that the optimal solution of eq.~(\ref{equ:ss4}) is
simply
\begin{align}
\widetilde{\bm{x}} = \bm{V} \widetilde{\bm{x}}_2 = \bm{V}
\begin{pmatrix}
    \bm{D}_1^{-1}\bm{d}_1 \\ \bm{0}
\end{pmatrix},
\end{align}
which is equivalent to the following:
\begin{align}\label{equ:ss6}
\widetilde{\bm{x}} = \bm{V}
\begin{pmatrix}
    \bm{D}_1^{-1} & \bm{0} \\ \bm{0} & \bm{0}
\end{pmatrix}\bm{U}^t \bm{d}
\end{align}

In fact, eq.~(\ref{equ:ss4}--\ref{equ:ss6}) are mainly used to derive
the solution. Once we obtain eq.~(\ref{equ:ss6}), there is no problem to
proceed to the final form of the solution that is free from zero rows or
columns. Because both $\bm{U}$ and $\bm{V}$ can be divided in columns as
\begin{align}
\bm{U} = \begin{pmatrix}
\bm{U}_1 & \bm{*}
\end{pmatrix}, \,
\bm{V} = \begin{pmatrix}
\bm{V}_1 & \bm{*}
\end{pmatrix};
\end{align}
thus,
\begin{align}\label{equ:ss7}
\widetilde{\bm{x}} =
\begin{pmatrix}
\bm{V}_1 & \bm{*}
\end{pmatrix} \begin{pmatrix}
    \bm{D}_1^{-1} & \bm{0} \\ \bm{0} & \bm{0}
\end{pmatrix}
\begin{pmatrix}
\bm{U}_1^t \\ \bm{*}
\end{pmatrix} 
\bm{d} = \bm{V}_1 \bm{D}_1^{-1} \bm{U}_1^t \bm{d}.
\end{align}
The above equation gives the optimal solution of map-making, provided
that the covariance matrix singularity has already been solved by the
general approach in section~\ref{sub:deal with singular cov}.

\subsection{Features of the optimal solution}

Provided we have no prior knowledge of the missing modes of the input
sky map\footnote{The term ``no prior knowledge'' means the missing modes
are regarded as lost forever, and there is no additional knowledge or
information that can help to recover them.}, the optimal solution is the
unique solution that satisfies the three important features listed
below:
\begin{enumerate}
    \item \label{itm:ss1} Lossless: $\widetilde{\bm{x}}$ is lossless
    except for the modes that are lost before any solution (prior loss).
    
    \item \label{itm:ss2} Minimum error: The TOD-domain estimation error
    $\bm{\delta} = \bm{d}-\bm{A}\widetilde{\bm{x}}$ is minimized.
    
    \item Minimum length: Of all solutions that satisfy
    items~\ref{itm:ss1}--\ref{itm:ss2}, $\widetilde{\bm{x}}$ (ignore
    noise) has the lowest pixel domain length (or power).
\end{enumerate}
Therefore, $\widetilde{\bm{x}}$ is the best blind solution (BBL) of the
problem, where ``blind'' explicitly refers to the pre-condition that we
have no prior knowledge of the missing modes of the sky map.

In a less complicated scenario where singular noise covariance matrices
do not need to be fully addressed, a solution analogous to the above was
described by \citep{2017A&A...600A..60P}. Nonetheless, their methodology
relied on the employment of an eigenvalue decomposition, which may not
be applicable to non-square matrices and could potentially falter when
dealing with certain square matrices, such as those derived from the
Jordan normal form. Moreover, the eigen-problem of a real-valued system
commonly necessitates resolution within the complex space. Meanwhile, it
is also possible to deal with the singularity problem by means of
limits, like the one shown in~\citep{1996astro.ph.12006W}, which is
often implemented by the use of an additional small matrix; however,
this will inevitably propagate the side effect to all modes, which will
not happen in the above general solution. Even if for the simplified
solution in section~\ref{subsub:modified pseudo inverse}, the side
effect will be strictly limited within the missing modes.

\subsection{Possible simplifications}
\label{sub: simp}

In this section, we discuss possible simplifications to the general
optimal solution.

\subsubsection{With tightly associated singularities}
\label{subsub: simp I}

As shown in section~\ref{sub:deal with singular cov}, it is quite
complicated to deal with a singular covariance matrix in general.
However, in some cases, the singularity of the covariance matrix comes
from the posterior processing of TOD, i.e., the singularities of matrix
$\bm{A}$ and $\bm{C}$ are tightly associated, which can make the problem
easier. For example, assume both $\bm{A}^t\bm{A}$ and $\bm{C}$ are
non-singular at the beginning, but the TOD is later processed as
\begin{align}
\bm{d} \rightarrow \bm{M} \bm{d} = \bm{M}\bm{A}\bm{x} + \bm{M} \bm{n},
\end{align}
where $\bm{M}$ is singular. One can prove that in this case, the lower
parts of eq.~(\ref{equ:adfaklj}) are all zero and can be safely ignored
to leave only the upper half
\begin{align}
\bm{P}^t_1\bm{d} = \bm{P}^t_1 \bm{A} \bm{x} + \bm{\lambda}_1 \bm{n}_1
\;\; \Longrightarrow \;\; 
\bm{d}_1 = \bm{A}_1 \bm{x} + \bm{\lambda}_1 \bm{n}_1,
\end{align}
whose the noise covariance matrix is non-singular. 

In the above mentioned \emph{special case}, one can prove that the
optimal solution is similar to eq.(\ref{equ:solution simplest case}),
just replacing each matrix inversion with the pseudo inverse:
\begin{align}\label{equ:adsfhak9729823}
\widetilde{\bm{x}} =
(\bm{A}^t\bm{C}^{+}\bm{A})^{+}\bm{A}^t\bm{C}^{+}\bm{d}.
\end{align}

It should be noted that even if the singularity arises solely from
filtering, the matrix $\bm{A}$ and $\bm{C}$ are not always associated.
This is especially true if the filtering involves a subtractive
operation, such as
\begin{align}
    \bm{d} \rightarrow \bm{d}' = \bm{A}\bm{x} + \bm{n} - \sum_i \bm{T}_i,
\end{align}
Here, $\bm{T}_i$ refers to the external templates that are used to
eliminate or mitigate contaminations. Estimating the impact of
subtraction on the noise and signal covariance matrices is challenging
because their cross-correlation is often unpredictable, and especially,
for a blind estimate, the signal is assumed to be unknown. However, it
is necessary to consider the possibility that the noise and signal
covariance matrices may be affected differently.

\subsubsection{The application of modified pseudo inverse}
\label{subsub:modified pseudo inverse}

Furthermore, one can also use the modified pseudo inverse to deal with
a singular noise covariance matrix, which can also significantly
simplify the processes but does not require the singularity association
between $\bm{A}$ and $\bm{C}$.

The modified pseudo inverse is defined as follows: Let $\lambda_i$ be
the diagonal elements of a diagonal matrix $\bm{\lambda}$, the modified
pseudo inverse of $\bm{\lambda}$ is another diagonal matrix
$\bm{\lambda}^\times$ of the same shape, whose diagonal elements
$\lambda_i^\times$ are given by
\begin{align}
\lambda_i^\times=\left\{
\begin{matrix}
\lambda_i^{-1} & (\lambda_i \ne 0)\\
1 & (\lambda_i = 0)\\
\end{matrix}\right.
\end{align}
Apparently, if we change the above definition to $\lambda_i^\times=0$
when $\lambda_i = 0$, then the definition returns to the pseudo inverse,
usually noted as $\bm{\lambda}^+$. For a general non-diagonal matrix
$\bm{M}$, the pseudo and modified pseudo inverses are defined based on
its singular-value decomposition: Assume $\bm{M}$'s singular value
decomposition is $\bm{M} = \bm{V}\bm{\lambda}\bm{U}^t$, and then its
pseudo and modified pseudo inverses are:
\begin{align}
\bm{M}^+ &= \bm{V}\bm{\lambda}^+\bm{U}^t \\ \nonumber
\bm{M}^\times &= \bm{V}\bm{\lambda}^\times\bm{U}^t,
\end{align}
respectively. The main difference between $\bm{\lambda}^\times$ and
$\bm{\lambda}^+$ is that the modified pseudo inverse
$\bm{\lambda}^\times$ is full ranked (invertible); thus, it is always
safe to left-multiply $\bm{\lambda}^\times$ to both sides of
eq.~(\ref{equ:a1}) to get
\begin{align}
\bm{\lambda}^\times \bm{P}^t \bm{d} &= \bm{\lambda}^\times\bm{P}^t\bm{A} 
\bm{x} +
\begin{pmatrix}
\bm{n}_1 \\ \bm{0}
\end{pmatrix}.
\end{align}
Then we replace the noise term in the above equation with $\bm{n}_0$ for
approximation, and get
\begin{align}\label{equ:asdfadfw123}
\bm{\lambda}^\times \bm{P}^t \bm{d} &\approx 
\bm{\lambda}^\times \bm{P}^t\bm{A} \bm{x} +
\bm{n}_0,
\end{align} 
which eliminates the covariance matrix singularity at the price of
virtually injecting the following difference into the linear system:
\begin{align}
\begin{pmatrix}
\bm{n}_1 \\ \bm{0}
\end{pmatrix}  - \bm{n}_0 = \begin{pmatrix}
\bm{0} \\ \bm{\delta}
\end{pmatrix},
\end{align}
where $\bm{\delta}$ is a standard white noise with $\ell$ rows.
Therefore, the injection is strictly limited within the singular part,
and the energy of the virtually injected noise is relatively
$\ell/(n_{\rm tod}-\ell)$. Because $\ell \ll n_{\rm tod}$, the noise
injection is negligible in the vast majority of cases, and will not
significantly deteriorate the solution. Especially, because the noise
injection is only ``virtually'' (no actual injection), the side effect
is further reduced, making the corresponding solution near optimal.

When the modified pseudo inverse is adopted, the near-optimal solution
can be shown as
\begin{align}\label{equ:afaiusdf878321}
\widetilde{\bm{x}} =
(\bm{A}^t\bm{C}^{\times}\bm{A})^{+}\bm{A}^t\bm{C}^{\times}\bm{d},
\end{align}
which is similar to eq.~(\ref{equ:adsfhak9729823}), only that the
inversion of the \emph{covariance matrix} is given by the modified
pseudo inverse rather than the standard pseudo inverse.

\subsection{Summary of the optimal and recommended solutions}
\label{sub:summary of two solutions}

As described above, the true optimal solution should be given by the
following approach:
\begin{enumerate}
\item First appropriately deal with the covariance matrix singularity as
shown in section~\ref{sub:deal with singular cov}.
\item Then follow sections~\ref{sub:reducing the size} -- \ref{sub: main
matrix singularity} to solve the rest of the problem.
\end{enumerate}

Because the true optimal solution is quite complex, two simplified
solutions are given above to make life easier. The solution with pseudo
inverse (section~\ref{subsub: simp I}) requires a special precondition
to be valid, and will also cause more problem regarding the scan
strategy (to be discussed in section~\ref{sub:regarding the main scan}).
The near-optimal solution with modified pseudo inverse
(section~\ref{subsub:modified pseudo inverse}) required no precondition,
has strictly limited side effect, and is more robust. Therefore, the
near-optimal solution in eq. (\ref{equ:afaiusdf878321}) is recommended.

Compared with the well known minimum variance solution of the simplest
case in eq.~(\ref{equ:solution simplest case}), the recommended solution
in eq.~(\ref{equ:afaiusdf878321}) inverts the noise covariance matrix
$\bm{C}$ by the modified pseudo inverse, and other matrices by the
standard pseudo inverse. As discussed in section~\ref{subsub:modified
pseudo inverse}, the side effect of eq.~(\ref{equ:afaiusdf878321}) is
strictly limited in the singular part. Thus the difference to the true
optimal solution is almost always negligible.

Another advantage of eq.~(\ref{equ:afaiusdf878321}) is that it is
suitable for fast computation, for several reasons:
\begin{enumerate}
\item There is no need to compute the SVD of the huge matrix $\bm{A}$.

\item The inversion of $\bm{A}^t\bm{C}^\times\bm{A}$ is for an
$n_{\rm{pix}}\times n_{\rm{pix}}$ matrix, not an $n_{\rm tod}\times
n_{\rm tod}$ matrix, which is much easier.

\item Both $\bm{A}^t\bm{C}^\times\bm{A}$ and the covariance matrix are
symmetric, which is easier to handle.

\item The inversion of the noise covariance matrix $\bm{C}$ can be done
by e.g., the Fast Fourier transform, which is much faster than matrix
multiplication.

\item Because eq.~(\ref{equ:afaiusdf878321}) prevents the information
loss on the CMB signal as much as possible, it does not cause any
unrecoverable E-to-B leakage by itself, which is an important advantage
for detecting the primordial gravitational waves via CMB.
\end{enumerate}

We also point out an interesting fact about the near-optimal solution:
According to section~\ref{sub:deal with singular cov} and especially
eq.~(\ref{equ:092384alk}), if $\ell$ (the number of zero eigenvalue of
$\bm{C}$) is sufficiently big, then eq.~(\ref{equ:092384alk}) may
contain enough noiseless rows to solve $\bm{x}$ completely and
precisely, making all rest of the equations useless. The true optimal
solution can properly deal with this case, hence it is definitely better
than other solutions. However, in most cases, $\ell$ must be very small,
which makes the difference between the optimal and near-optimal solution
negligible. This is the precondition for eq.~(\ref{equ:afaiusdf878321})
to be useful.

\subsection{A quick reference}

This section is only a quick reference for busy readers, without any new
content.

For the problem of solving $\bm{x}$ from $\bm{d}=\bm{A}\bm{x}+\bm{n}$,
the optimal solution is given as follows: First follow the general
approach in section~\ref{sub:deal with singular cov} to deal with the
singularity of the covariance matrix and convert the noise to standard
white noise to obtain a \emph{renewed} main equation:
\begin{align}
\bm{d} = \bm{A} \bm{x} + \bm{n}_0,
\end{align}
which still allows a singular $\bm{A}^t\bm{A}$. The optimal solution of
the above \emph{renewed} equation is precisely:
\begin{align}
\widetilde{\bm{x}} = (\bm{A}^t\bm{A})^+ \bm{A}^t \bm{d}.
\end{align}

Practically, if the singularity of the covariance matrix is handled by
the modified pseudo inverse, then we get the near-optimal solution
directly from the \emph{original} linear system $\bm{d} = \bm{A}\bm{x}
+\bm{n} $ as
\begin{align}\label{equ:923ihadf}
\widetilde{\bm{x}} \approx
(\bm{A}^t\bm{C}^\times\bm{A})^+\bm{A}^t\bm{C}^\times\bm{d}.
\end{align}

Although the solution in eq.~(\ref{equ:923ihadf}) may be slightly
sub-optimal, it is recommended because it involves a considerably easier
computational procedure than the optimal option. Additionally, the
divergence from the optimal solution is strictly limited to the singular
components, which is usually insignificant.

In the list below, we compare the sub-optimal solution $\bm{S}_\times$
and the similar solution that uses only the pseudo-inverse, $\bm{S}_+$:
\begin{enumerate}
    \item When $\bm{A}^t\bm{A}$ and $\bm{C}$ are both non-singular: Both
    $\bm{S}_\times$ and $\bm{S}_+$ are optimal, and the signal is
    lossless in the solution.
    \item When $\bm{A}^t\bm{A}$ is singular but $\bm{C}$ is
    non-singular: $\bm{S}_\times$ is slightly sub-optimal, $\bm{S}_+$ is
    optimal. However, none of them can keep the signal lossless.
    \item When $\bm{A}^t\bm{A}$ is non-singular but $\bm{C}$ is
    singular: $\bm{S}_\times$ is near optimal with strictly limited
    negative effect, and keeps the signal lossless; $\bm{S}_+$ is non-
    optimal and will hurt the signal. $\bm{S}_\times$ is hence almost
    always better than $\bm{S}_+$.
    \item When $\bm{A}^t\bm{A}$ and $\bm{C}$ are both singular: Can be
    one of above three.
\end{enumerate} 
In all cases, the side effect of $\bm{S}_\times$ is strictly limited,
and can keeps the signal lossless as much as possible. This is why we
tends to adopt $\bm{S}_\times$ as the recommended solution.

However, we need to point out the following fact: as mentioned in
section~\ref{sub:regarding the main scan}, when the mapmaking algorithm
is designed to keep the signal lossless (as a precondition), there is
going to be a significant amplification of the noise term in the modes
with poor SNR, and the overall mapmaking result may look bad. This is
an inevitable consequence of all ``lossless'' or ``make lossless at all
costs'' mapmaking algorithms. The solution is also simple: one needs to
consider a posterior Wiener filtering style approach to put a reasonable
balance between the signal loss and noise control.

\section{Efficient computation of the noise covariance matrix's modified
pseudo inverse}

To calculate the sub-optimal solution outlined in equation
(\ref{equ:923ihadf}), it is necessary to determine the modified
pseudo-inverse of matrix $\bm{C}$. However, it should be noted that
$\bm{C}$ is a matrix in the time domain, which may have a significantly
larger size compared to a pixel domain matrix. Consequently, it becomes
imperative to explore efficient methods for computing $\bm{C}^\times$.

Usually, $\bm{C}$ can be regarded as diagonal in the Fourier domain (at
least approximately), so its time-domain pseudo inverse is simply the
following:
\begin{align}
\bm{C} = \bm{W}\bm{\lambda}^2\bm{W}^H \longrightarrow 
\bm{C}^\times = \bm{W}(\bm{\lambda}^2)^\times\bm{W}^H,
\end{align}
where $\bm{W}$ is the Fourier transform matrix. However, when the noise
is filtered like $\bm{n} \longrightarrow \bm{M} \bm{n}$, the above
equation can no longer be used to compute the modified pseudo inverse of
$\bm{C}$. In this case, we first consider the singular value
decomposition of the filtering matrix as $\bm{M} =
\bm{U}\bm{\Lambda}\bm{V}^H$, and convert the time domain noise to white
noise as shown in eq.~(\ref{equ:a1}), $\bm{n} = \bm{P} \bm{\lambda}
\bm{n}_0$, to obtain
\begin{align}
\bm{d} = \bm{A}\bm{x} + \bm{M}\bm{n} = 
\bm{A}\bm{x} + \bm{U}\bm{\Lambda}\bm{V}^H\bm{P}\bm{\lambda}\bm{n}_0.
\end{align}
Left-multiply $\bm{U}^H$ to both sides:
\begin{align}
\bm{U}^H\bm{d} = 
\bm{U}^H\bm{A}\bm{x} + \bm{\Lambda}\bm{V}^H\bm{P}\bm{\lambda}\bm{n}_0,
\end{align}
then with the idea of modified pseudo inverse, left-multiply
$\bm{\Lambda}^\times$ to both sides:
\begin{align}
\bm{\Lambda}^\times\bm{U}^H\bm{d} = 
\bm{\Lambda}^\times\bm{U}^H\bm{A}\bm{x} + 
\bm{\Lambda}^\times\bm{\Lambda}\bm{V}^H\bm{P}\bm{\lambda}\bm{n}_0,
\end{align}
and assume $\bm{\Lambda}^\times\bm{\Lambda}\approx\bm{I}$. As explained
in section~\ref{subsub:modified pseudo inverse}, this means to assume
some additional noise in the singular part, whose side effect is
strictly limited and usually insignificant. With the approximation, we
further get
\begin{align}
\bm{\lambda}^\times\bm{P}^H\bm{V}\bm{\Lambda}^\times\bm{U}^H\bm{d} \approx 
\bm{\lambda}^\times\bm{P}^H\bm{V}\bm{\Lambda}^\times\bm{U}^H\bm{A}\bm{x} + 
\bm{n}_0,
\end{align}
whose solution is
\begin{align}\label{equ:fast pinv time domain1}
\widetilde{\bm{x}} = (\bm{A}_1^H\bm{A}_1)^+ \bm{A}_1^H \bm{d}_1,
\end{align}
which does not require to compute $\bm{C}^\times$, and
\begin{align}\label{equ:fast pinv time domain2}
\bm{A}_1 &= \bm{\lambda}^\times\left(\bm{P}^H\bm{V}\bm{\Lambda}^\times 
\bm{U}^H\right)\bm{A} = \bm{\lambda}^\times(\bm{P}^H\bm{M}^\times)\bm{A} \\ \nonumber
\bm{d}_1 &= \bm{\lambda}^\times\left(\bm{P}^H\bm{V}\bm{\Lambda}^\times 
\bm{U}^H\right)\bm{d} = \bm{\lambda}^\times(\bm{P}^H\bm{M}^\times)\bm{d}.
\end{align}
As long as the filtering scheme does not change, $\bm{M}$ is a constant
matrix, so are $\bm{M}^\times$ and $\bm{P}^H\bm{M}^\times$. Therefore,
with eqs.~(\ref{equ:fast pinv time domain1}--\ref{equ:fast pinv time
domain2}), the time domain eigen-problem is converted to matrix
multiplication, which can be computed efficiently  once the noise
spectrum $\bm{\lambda}$ is obtained.

\section{More discussion about the singularity of the main matrix}\label{appendix:C}

It is not always easy to determine the singularity of matrix
$\bm{A}^t\bm{C}^+\bm{A}$ when $\bm{C}^+$ is a singular matrix. To
determine the rank of matrix $\bm{A}^t\bm{C}^+\bm{A}$, we can find the
dimension of the null-space of it using the rank-nullity theorem.

If $\bm{v}\in {\rm Null}(\bm{A}^t\bm{C}^+\bm{A})$, then it implies that
\begin{align}
    \bm{v}^t\bm{A}^t\bm{C}^+\bm{A}\bm{v} = 0.
\end{align}
Assume that $\bm{A}$ has full column-rank, which means for any
$\bm{v}\neq\bm{0}$, $\bm{Av}\neq \bm{0}$. Thus $\bm{Av} \in {\rm
Null}(\bm{C}^+)$. Due to the fact that $\bm{C}^+$ and $\bm{C}$ have the
same null-space because $\bm{C}$ is symmetric, it is evident that
\begin{align}
    \bm{v}^t\bm{A}^t\bm{C}\bm{A}\bm{v} = 0.
\end{align}
Therefore, we can conclude that ${\rm
Null}(\bm{A}^t\bm{C}^+\bm{A})\subseteq{\rm Null}(\bm{A}^t\bm{C}\bm{A})$.
On the contrary, assuming that $\bm{v} \in {\rm
Null}(\bm{A}^t\bm{C}\bm{A})$, we can also conclude that ${\rm
Null}(\bm{A}^t\bm{C}\bm{A})\subseteq{\rm Null}(\bm{A}^t\bm{C}^+\bm{A})$.
These two conclusions indicate that $\bm{A}^t\bm{C}^+\bm{A}$ and
$\bm{A}^t\bm{C}\bm{A}$ have the same null-space.

Usually, the singularity of matrix $\bm{C}$ is caused by filtering.
When applying a filter to TOD, the covariance matrix will be modified,
and this modification can be represented as $\bm{C} =
\bm{M}\tilde{\bm{C}}\bm{M}^t$, where $\tilde{\bm{C}}$ is a
positive-definite matrix,  and $\bm{M}$ represents the filter. Again, if
$\bm{v} \in {\rm Null}(\bm{A}^t\bm{C}\bm{A})$, we have
\begin{align}
    \bm{v}^t\bm{A}^t\bm{M}\tilde{\bm{C}}\bm{M}^t\bm{A}\bm{v} = 0.
\end{align}
Since $\tilde{C}$ is positive-definite, we can state that the equation
holds if and only if $\bm{M}^t\bm{Av}=\bm{0}$, which also means that
$\bm{A}^t\bm{C}\bm{A}$ and $\bm{M}^t\bm{A}$ have the same null-space.
Thus $\bm{A}^t\bm{C}^+\bm{A}$ and $\bm{M}^t\bm{A}$ have the same
null-space.

Commonly used filters, such as high-pass, polynomial, and Wiener
filters, are symmetric, meaning that $\bm{M}=\bm{M}^t$. We can write SVD
of $\bm{M}$ and $\bm{A}$ as \begin{align}
    \bm{M} &= \bm{U}\bm{\lambda}\bm{V}^t &  \bm{A} &= \bm{P}\bm{\delta}\bm{Q}^t.   
\end{align}
Thus $\bm{MA} = \bm{U\lambda V}^t\bm{P\delta Q}^t$. Let $\bm{G}$ be
represented as $\bm{V}^t\bm{P}$. It is evident that $\bm{G}$ is an
$n_{\rm tod}$-by-$n_{\rm tod}$ unitary matrix. Divide $\bm{\lambda}$, $\bm{G}$
and $\bm{\delta}$ into blocks as
\begin{align}
    \bm{\lambda} &= \begin{pmatrix}
        \bm{\lambda}_1 & \bm{0} \\
        \bm{0} & \bm{0}
    \end{pmatrix} &  \bm{G} &=\begin{pmatrix}
        \bm{G}_1 & \bm{G}_2 \\
        \bm{G}_3 & \bm{G}_4
    \end{pmatrix} & \bm{\delta} &= \begin{pmatrix}
        \bm{\delta}_1 \\ 
        \bm{0}
    \end{pmatrix},
\end{align}
then the only singular part in $\bm{MA}$ can be denoted as 
\begin{align}
    \bm{\lambda G \delta} = \begin{pmatrix}
        \bm{\lambda}_1 \bm{G}_1 \bm{\delta}_1 & \bm{0}\\
        \bm{0} & \bm{0}
    \end{pmatrix},
\end{align} 
which means the final rank is determined by the rank of the upper left
corner of $\bm{G}$. The upper left part of $\bm{G}$ is formed by
multiplying the eigenvector matrix of $\bm{M}$ with the column singular
vectors of $\bm{A}$, which is equivalent to performing a Fourier
transform on the column singular vectors of $\bm{A}$. Therefore, the
final rank depends on the Fourier space structure of the column singular
vectors of $\bm{A}$. If their main differences in Fourier space happen
to be concentrated in the zeroed-out lower-left part, singularity will
occur. For random scanning, the Fourier space structure of the column
singular vectors of $\bm{A}$ should be randomly distributed among
different components, making it unlikely to be concentrated in the
lower-left part. Hence, singularity is hard to appear. However, there is
one exception when $\bm{v}$ is proportional to $\bm{1}_{n_{\rm pix}}$.
In this case, $\bm{Av}$ is proportional to $\bm{1}_{n_{\rm tod}}$, and
if the filter $\bm{M}$ removes the mean value, then $\bm{MAv} = 0$.

In summary, if the filter matrix $\bm{M}$ is capable of filtering out
the monopole component, the rank of $\bm{A}^t\bm{C}^+\bm{A}$ is no more
than $n_{\rm pix}- 1$. However, if the filter matrix $\bm{M}$ does not
remove the monopole component, the rank can be $n_{\rm pix}$ in the best
case.

\section{The code validation}
\label{appendix:validation}

All three solutions $\bm{S}_+$, $\bm{S}_\times$, and $\bm{S}_{\rm c}$
are unbiased, with the exception of the lost modes. In computing
$\bm{C}^{+}$, we set a threshold and ignore the small singular values in
order to improve the numerical stability. Assuming that the suppression
effect is independent at each $\ell$, the final power spectra can be
corrected by running several simulations to get the suppression factor.
Then we can evaluate the effectiveness of the three solutions
$\bm{S}_\times, \bm{S}_+$, and $\bm{S}_{\rm c}$ by comparing their
uncertainties because they are all unbiased for the remaining modes.

\begin{figure}[!ht]
    \centering
    \includegraphics[width=0.45\textwidth]{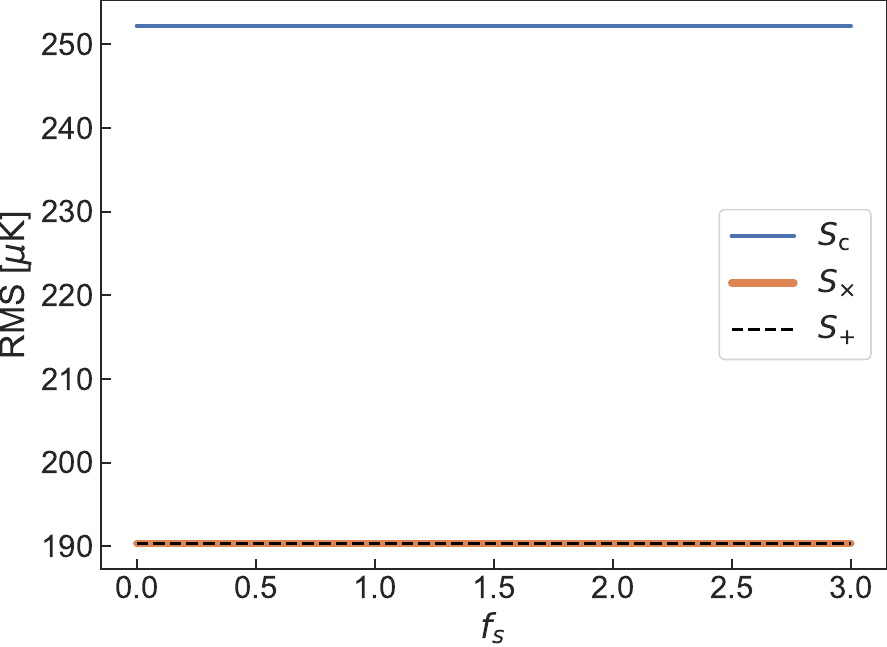}
    \includegraphics[width=0.45\textwidth]{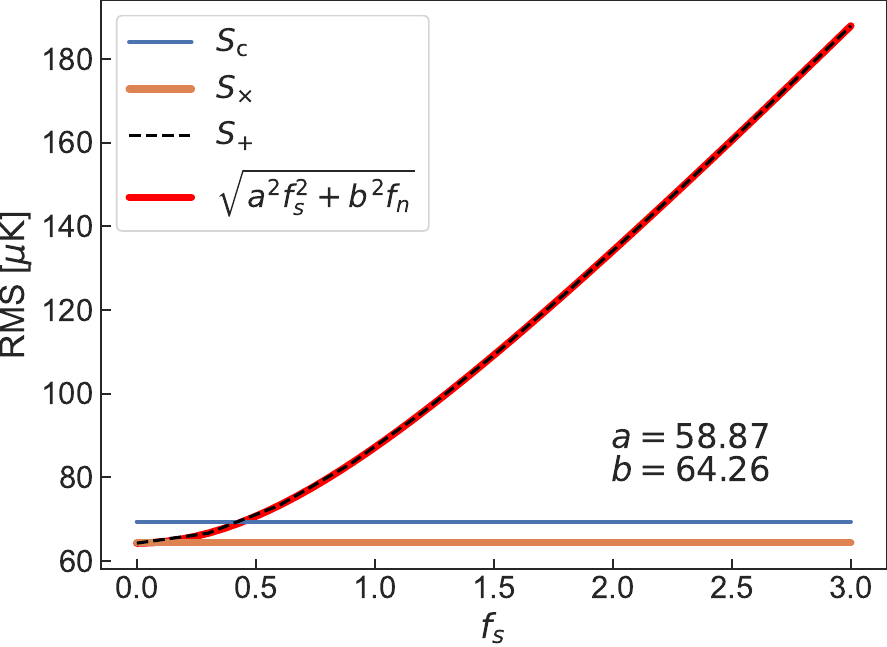}
    \caption{The pixel domain RMS (Root Mean Square) of the residual
    (output-input) as a function of $f_s$, with $f_n$ fixed at 1 for a
    non-singular covariance matrix (left), and a singular covariance
    matrix (right).}
    \label{fig:change fs}
\end{figure}

In order to validate the properties of $\bm{S}_+,\bm{S}_\times$ and
$\bm{S}_{\rm c}$, the magnitudes of the CMB signal and total noise are
modulated as $\bm{d} = f_s \bm{d}_{\rm cmb} + f_n\bm{n}$. In the context
of a solution that enables complete reinstatement of the CMB signal
(lossless), the solution error is determined only by the noise.
Consequently, upon fixing the noise amplitude $f_n$ and varying only the
CMB amplitude $f_s$ during the simulation, the ensuing errors in the
$\bm{S}_{\rm c}$ and $S_{\rm \times}$ solutions should not change with
$f_s$. This holds true regardless of the singularity of the noise
covariance matrix $\bm{C}$. However, there should arises a difference in
the case of $\bm{S}_{\rm +}$: If the noise covariance matrix $\bm{C}$ is
indeed singular, the error in $\bm{S}_{\rm +}$ will exhibit a change
with $f_s$, because under these specific conditions, $\bm{S}_{\rm +}$
leads to a loss in the CMB signal, and the residual RMS will hence scale
as $\sqrt{a^2 f_s^2+b^2 f_n^2}$. All these expectations are confirmed by
the data depicted in Figure~\ref{fig:change fs}.

\begin{figure}[ht]
    \centering
    \includegraphics[width=0.45\textwidth]{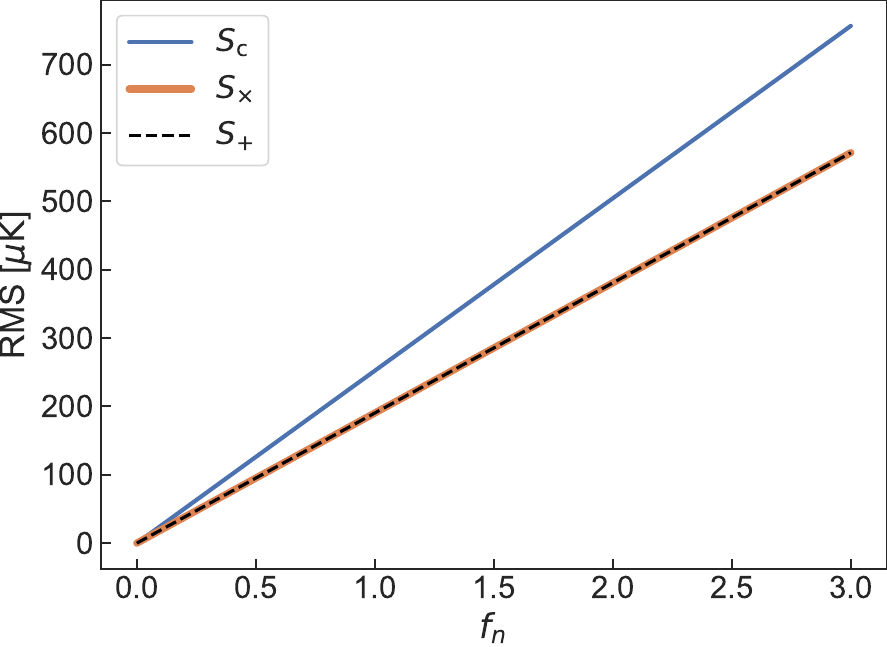}
    \includegraphics[width=0.45\textwidth]{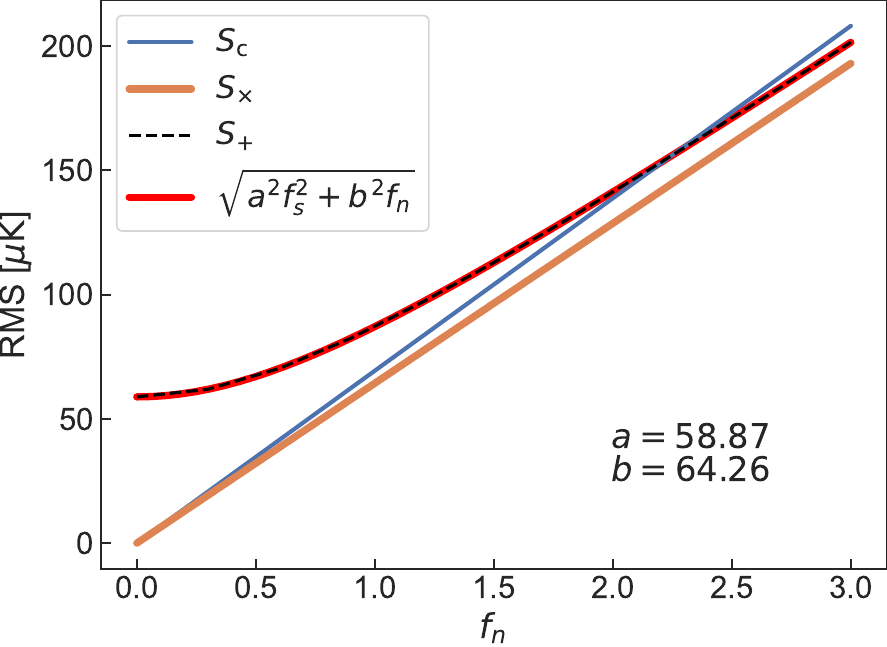}
    \caption{Similar to figure~\ref{fig:change fs} but change $f_n$
    instead.}\label{fig:change fn}
\end{figure}

Alternatively, if one maintains a constant CMB amplitude and varies the
noise amplitude $f_n$ during simulation, the errors in $\bm{S}_{\rm c}$
and $S_{\rm \times}$ should reveal a direct proportionality to $f_n$
because there is no CMB error in them (lossless), which applies
regardless of the singularity of the noise covariance matrix $\bm{C}$.
In contrast, when the noise covariance matrix $\bm{C}$ is singular, the
error in $\bm{S}_{\rm +}$ should not exhibit a linear relationship with
$f_n$, because it contains contributions both from the lost CMB signal
and the residual noise. This phenomenon is also confirmed by the data
presented in Figure~\ref{fig:change fn}. Furthermore, both
Figure~\ref{fig:change fs}--\ref{fig:change fn} show that $S_{\rm
\times}$ has a significantly lower residual noise than $\bm{S}_{\rm c}$,
thereby corroborating its superior efficacy in red noise suppression.

The results in Figure~\ref{fig:change fs}--\ref{fig:change fn} are well
consistent with expectations, which confirms the validity of our code
implementation and ensures reliable results in the tests that follows.


\providecommand{\href}[2]{#2}\begingroup\raggedright\endgroup

\end{document}